\documentclass[journal, 10pt]{IEEEtran}

\usepackage{cite}
\usepackage{amsmath,amssymb,amsfonts}
\usepackage{algorithmic}
\usepackage{graphicx}
\usepackage{subfigure}
\usepackage{textcomp}
\usepackage{xcolor}
\usepackage{url}
\usepackage{bbm}
\usepackage{bm}
\usepackage[ruled, linesnumbered]{algorithm2e}
\allowdisplaybreaks[4]

\ifCLASSINFOpdf
\else
\fi
\begin{document}

\title{DFT-Spread Orthogonal Time Frequency Space System with Superimposed Pilots for Terahertz Integrated Sensing and Communication}
%

\author{Yongzhi~Wu,~\IEEEmembership{Graduate~Student~Member,~IEEE,}
        Chong~Han,~\IEEEmembership{Member,~IEEE,}
        and~Zhi~Chen,~\IEEEmembership{Senior Member,~IEEE}
\thanks{
This work was presented in part at IEEE International Conference on Communications, 2022~\cite{wu2022dftsotfs-isac}. This work was in-part supported by National Key R\&D Program of China under Project No. 2020YFB1805700, and ``the Fundamental Research Funds for the Central Universities".

Yongzhi~Wu and Chong~Han are with the Terahertz Wireless Communications (TWC) Laboratory, Shanghai Jiao Tong University, Shanghai, China (Email:~\{yongzhi.wu,~chong.han\}@sjtu.edu.cn). Chong Han is also affiliated with Department of Electronic Engineering and Cooperative Medianet Innovation Center (CMIC), Shanghai Jiao Tong University.

Zhi Chen is with University of Electronic Science and Technology of China, Chengdu, China (Email:~chenzhi@uestc.edu.cn).}
}

\maketitle
\thispagestyle{empty}


\begin{abstract}
	Terahertz (THz) integrated sensing and communication (ISAC) is a promising interdisciplinary technology that realizes simultaneously transmitting Terabit-per-second (Tbps) and millimeter-level accurate environment or human activity sensing. However, both communication performance and sensing accuracy are influenced by the Doppler effects, which are especially severe in the THz band. Moreover, peak-to-average power ratio (PAPR) degrades the THz power amplifier (PA) efficiency. In this paper, a discrete Fourier transform spread orthogonal time frequency space (DFT-s-OTFS) system with superimposed pilots is proposed to improve the robustness to Doppler effects and reduce PAPR for THz ISAC. Then, a two-phase sensing parameter estimation algorithm is developed to integrate sensing functionality into the DFT-s-OTFS waveform.
	Meanwhile, a low-complexity iterative channel estimation and data detection method with a conjugate gradient based equalizer is proposed to recover the data symbols of DFT-s-OTFS. The proposed DFT-s-OTFS waveform can improve the PA efficiency by 10\% on average compared to OTFS. Simulation results demonstrate that the proposed two-phase sensing estimation algorithm for THz DFT-s-OTFS systems is able to realize millimeter-level range estimation accuracy and decimeter-per-second-level velocity estimation accuracy. Moreover, the effectiveness of the iterative method for data detection aided by superimposed pilots in DFT-s-OTFS systems is validated by the simulations and the bit error rate performance is not degraded by the Doppler effects.
\end{abstract}



\section{Introduction}

\IEEEPARstart{R}{ecently}, with the exhaustion of spectrum resource in the microwave band and rapid growth of wireless data rates, higher and wider spectrum is demanded to satisfy the key performance metrics of the sixth generation (6G) wireless systems. Along with the trend of moving up to higher frequencies, the Terahertz (THz) band (0.1-10 THz) is viewed as one of the key technologies to realize the promising 6G blueprint: all things are sensing, connected, and intelligent~\cite{chen2021THz}.

On one hand, wireless Terabit-per-second (Tbps) links will become a reality in the 2030 and beyond intelligent information society, thanks to the ultra-broad bandwidth in the THz band. On the other hand, ultra-accurate sensing, e.g., millimeter-level range estimation and millidegree-level angle estimation can be realized.  Furthermore, the system design aims at \textit{Terahertz integrated sensing and communication (THz ISAC)}, which opens up new applications, e.g., vehicle-to-vehicle networks and THz Internet-of-Things (Tera-IoT)~\cite{chen2021THz}. Moreover, by sharing the spectrum, hardware and waveform, THz ISAC can reduce the hardware cost, and improve the spectral and energy efficiency~\cite{liu2020jrc}.

While enabling ultra-accurate sensing and faster connections, THz ISAC encounters two critical challenges on the waveform design. First, towards higher carrier frequencies, the Doppler spread effect becomes severer in the THz band, especially in high-mobility scenarios~\cite{chen2019vehicle}. Fast time-varying channels with high Doppler spread may cause inter-carrier interference (ICI) in the time-frequency (TF) domain. Waveforms that work in the TF domain, such as orthogonal frequency division multiplexing (OFDM) and discrete Fourier transform (DFT) spread OFDM (DFT-s-OFDM, aka SC-FDMA), meet trouble in equalizing multiple Doppler shifts along each path. As a result, the link performance would be seriously deteriorated in terms of the bit error rate (BER) and data rate performance if maintaining current waveform and numerology. Meanwhile, sensing parameter estimation in OFDM radar receiver usually utilizes the signal model without considering the ICI, which causes estimation error in the presence of high-speed targets. In this case, the sensing accuracy using OFDM waveform would be degraded.

The second challenge is that with the huge amount of wireless devices, energy efficiency is expected to be improved by 100 times in 6G wireless systems. Nevertheless, when moving up to higher operational frequencies, the saturated output power of power amplifiers (PAs) rapid decreases in spite of the integrated circuit technology~\cite{wang2020pa}. In order to maximize the transmit power and energy efficiency of PAs in the THz band, the peak-to-average power ratio (PAPR) requirement on the transmit signal tends even stricter compared with the microwave band. Therefore, when designing the THz ISAC transmit waveform that is commonly used by sensing and communication, we need to address the aforementioned issues.

\subsection{Related Work}

In this section, we first investigate the ISAC systems in the microwave band and then elaborate that they are unable to satisfy requirements of THz ISAC.
In the literature, ISAC systems are implemented in different ways. In addition to some resource-sharing schemes, such as dividing time resources into either sensing or communication~\cite{petrov2019jcs}, existing works on integrated sensing and communication consider using a common waveform~\cite{zhang2021enabling}, which includes embedding message into radar waveforms and realizing sensing parameter estimation based on the standard communication waveforms. For example, authors in \cite{Strum2011Waveform} propose an OFDM ISAC framework with the two-dimensional (2D) fast Fourier transform (FFT) method and a multiple signal classification method (MUSIC). A correlation-based radar processing approach is used to estimate radar sensing parameters based on the OFDM baseband signals in~\cite{Berger2010ofdm}. Nevertheless, OFDM-based ISAC systems suffers from a high PAPR and thus is not energy-efficient, especially for THz systems. Authors in \cite{Kumari2018radar} exploit the preamble of a single-carrier physical
layer frame and make it suitable for sensing. Due to lower PAPR, single-carrier waveforms have advantages over multi-carrier modulations in terms of the energy-efficiency. However, both conventional single-carrier and multi-carrier waveforms meet challenges in Doppler effects when it comes to the THz band.

Along with the increase of carrier frequencies, the Doppler shift, which is proportional to the carrier frequency, becomes larger and thus causes stronger Doppler effect. For instance, the Doppler shift is enlarged by 100 times when the frequency is increased  from 3 GHz to 0.3 THz. The severe Doppler spread destroys the orthogonality of subcarriers in OFDM and cause ICI in the frequency domain, and thus, the Doppler shift are difficult to estimate and equalize in OFDM systems.
Recently, an orthogonal time frequency space (OTFS) modulation scheme is proposed to deal with high Doppler spread in doubly selective channels~\cite{raviteja2018otfs}. OTFS modulates information in the delay-Doppler (DD) domain and conveniently accommodates the channel dynamics, which shows significant advantages over OFDM in the presence of severe Doppler effects~\cite{wei2021otfs}. A time-varying multipath channel can be transformed into a near-stationary channel in the delay-Doppler domain~\cite{surabhi2019diversity}. Meanwhile, the effectiveness of OTFS for ISAC is validated in~\cite{gaudio2020otfs}. Nevertheless, the time domain transmit samples of OTFS are equivalent to the output of inverse DFT (IDFT) of the information symbols in the delay-Doppler domain~\cite{Surabhi2019otfs}, which behaves like OFDM with a relatively high PAPR value. The PAPR of OTFS is lower than OFDM but still not satisfactory for THz PAs. Thus, since OTFS has similar PAPR problem to OFDM, a PAPR reduction scheme is required to improve the THz PA efficiency.
In this paper, the DFT spreading operation is leveraged for OTFS to spread the input signal with DFT, which can effectively reduce the PAPR of OTFS signal by about 3 dB. This reduction makes the PAPR of the resulting DFT-s-OTFS at the same level of single-carrier transmission, while maintaining the robustness against severe Doppler effects in the THz band.

Till date, most studies working on OTFS modulation assume that the delay and Doppler shifts of channel impulse response are integer multiples of delay and Doppler resolution, respectively. While the delay resolution is sufficiently large for communication systems, the presence of fractional Doppler may cause channel spreading across the Doppler indices and breaks the channel sparsity in the delay-Doppler domain~\cite{wei2021otfs}. A Dolph-Chebyshev (DC) window design is proposed in~\cite{wei2021window} to suppress the channel spreading and improved the channel estimation accuracy.
Channel estimation and data detection with fractional delay and Doppler have also been investigated in some work, such as \cite{wei2022ce, thaj2020equalizer, yuan2022otfs}. A sparse Bayesian
learning method is proposed in~\cite{wei2022ce} to perform off-grid channel estimation for OTFS, while it applies an ideal pulse shaping filter. \cite{thaj2020equalizer} proposes an iterative rake decision feedback equalizer for the zero-padded OTFS system. An efficient unitary approximate message passing based detector is developed in~\cite{yuan2022otfs} for OTFS with fractional Doppler, but it appends a cyclic prefix (CP) for each symbol when using rectangular pulse.
To achieve high-accuracy sensing in THz ISAC systems, the channel delay and Doppler parameters should not be limited to integer values of delay and Doppler resolution.
Although the OTFS channel matrix for continuous-valued delay and Doppler shift is derived in~\cite{gaudio2020otfs}, the derived result is obtained by using the assumption of rectangular transmit pulse and the approximation of the integral with a discrete sum. Hence, an exact channel matrix model for continuous-valued delay and Doppler is still needed for THz ISAC systems.

Despite the promising communication and sensing abilities, the implementation complexity of OTFS is a pivotal issue, specifically, the complexity of OTFS channel estimation and data detection. The requirement of computational complexity is more stringent in the THz band to realize high-speed baseband digital processing. Low complexity channel estimation and detection of OTFS signals have been studied in \cite{raviteja2018otfs, raviteja2019pilot, mishra2021otfs, yuan2021sp, li2021otfs, liu2020mimo-otfs, liu2021otfs, yuan2020otfs, tiwari2019otfs, surabhi2020otfs, singh2022mimo-otfs, singh2022otfs, qu2021otfs}. An embedded pilot (EP) based OTFS frame structure is proposed in~\cite{raviteja2019pilot}, where the guard symbols are arranged to preserve the pilot symbol from the interference data symbols. In this case, the insertion of guard symbols results in non-negligible DD resource overhead and reduces the spectral efficiency. Schemes of superimposed pilot design are developed in~\cite{mishra2021otfs, yuan2021sp}, while \cite{mishra2021otfs} still uses EP frame to estimate delay and Doppler parameters and \cite{yuan2021sp} does not consider fractional delay and Doppler. Channel estimation approaches for MIMO-OTFS system are developed in~\cite{li2021otfs, liu2020mimo-otfs}, while they add cyclic prefix for each symbol and only consider integer delay. Message passing based methods are proposed to estimate channel parameters and design a data detector in~\cite{raviteja2019pilot, liu2021otfs}. A variational Bayes approach is developed in~\cite{yuan2020otfs} to reduce the receiver complexity of the OTFS receiver. Nevertheless, when the information symbols are transformed with a DFT precoding rather than being directly mapped on the DD domain,
these maximum a posterior probability (MAP) detection approaches can not exploit the sparsity of the delay-Doppler domain channel, which influences the computational complexity.
Low complexity minimum mean square error (MMSE) channel equalizers for OTFS are proposed under the assumption of integer delay and Doppler in the literature, which exploit the quasi banded structure of matrices for a practical rectangular pulse~\cite{tiwari2019otfs} and the block-circulant property of channel matrix for an ideal bi-orthogonal pulse~\cite{surabhi2020otfs}, but these properties becomes invalid when considering non-integer delay and Doppler. By utilizing the block-circulant structure, low complexity zero-forcing (ZF) and MMSE receivers have been developed for MIMO-OTFS systems in~\cite{singh2022mimo-otfs, singh2022otfs}.
In summary, to satisfy the requirements of THz ISAC, OTFS needs to overcome the problems, including PAPR, pilot design with reduced overhead, low-complexity channel estimation and data detection with fractional delay and Doppler.

In light of the aforementioned challenges of THz ISAC, the THz waveform needs to be well designed in terms of the PA efficiency, bit error rate (BER), robustness to Doppler effects, sensing accuracy and implementation complexity, which motivates this study. In our preliminary and shorter version~\cite{wu2022dftsotfs-isac}, we propose a sensing parameter estimation method by employing the data signals of DFT-s-OTFS. In this paper, we further design a sensing integrated DFT-spread OTFS (SI-DFT-s-OTFS) system with superimposed pilots, and develop low-complexity sensing parameter estimation and data detection approaches. In contrast with existing works on OTFS with superimposed pilots~\cite{mishra2021otfs, yuan2021sp}, which only consider integer delay and Doppler, we propose a high-accuracy estimation method for fractional parameters. Moreover, message passing based data detector and MAP symbol detection are respectively used in~\cite{mishra2021otfs} and~\cite{yuan2021sp},
but they are not able to make use of channel sparsity in the delay-Doppler domain for DFT-s-OTFS.

\subsection{Our Contributions}

The contributions of this work are summarized as follows:
\begin{itemize}
    \item \textbf{We propose a DFT-s-OTFS system framework for THz ISAC by developing a DFT precoding operation and designing a scheme of superimposed pilots in the delay-Doppler domain.} We design an optimal power allocation scheme between pilot and data by deriving the effective signal-to-interference-plus-noise (SINR) of received signal. The average pilot power that optimizes the BER performance depends on the signal-to-noise ratio (SNR), which is validated by the simulation results.
    \item \textbf{We propose a two-phase sensing parameter estimation algorithm for multiple target estimation in DFT-s-OTFS ISAC systems.} Meanwhile, to achieve super-resolution sensing accuracy, we also derive a continuous-delay-and-Doppler-shift (CDDS) channel matrix model for the developed sensing integrated DFT-s-OTFS systems. Based on the CDDS channel matrix, the two-phase estimation (TPE) method incorporates coarse on-grid search with low complexity in the first phase and refined off-grid search in the second phase.
    \item \textbf{We develop a low-complexity iterative channel estimation and data detection method with a conjugate gradient based equalizer for DFT-s-OTFS systems.} The proposed iterative method initializes the channel estimates aided by the pilot and then iteratively performs data detection and data-aided channel estimation to refine the estimation accuracy and detection performance. The developed conjugate gradient based channel equalization implements the computation complexity of $\mathcal{O}(MN \log(MN))$ by employing the partial Fourier form of CDDS channel matrix.
    \item \textbf{We conduct extensive simulation results of DFT-s-OTFS with superimposed pilots for THz ISAC.} The transmit signal of the proposed DFT-s-OTFS is able to improve the PA efficiency by 10\% compared with OTFS due to the reduction of PAPR. The simulation results validate the effectiveness of the proposed iterative method and the robustness of DFT-s-OTFS to strong Doppler effects. With the proposed TPE method, simulation results indicate that the THz SI-DFT-s-OTFS systems are able to realize millimeter-level range estimation accuracy and decimeter-per-second-level velocity estimation accuracy.
\end{itemize}

The structure of this paper is as follows. The system framework of DFT-s-OTFS for THz ISAC is elaborated in Sec.~\ref{sec:system}. The sensing parameter estimation algorithm is proposed in Sec.~\ref{sec:active_sensing}. In Sec.~\ref{sec:passive_sensing}, the iterative channel estimation and data detection is developed. Sec.~\ref{sec:power_allocation} delineates the optimal power allocation between pilot and data. Extensive simulation results are described in Sec.~\ref{sec:simulation}. Finally, Sec.~\ref{sec:conclusion} concludes the paper.

\textbf{Notations:} $\mathbb{C}$, $\mathbb{R}$ and $\mathbb{R}_{+}$ denote the set of complex numbers, real numbers and positive numbers, respectively. $\mathbb{E}\{\cdot\}$ defines the expectation operation. The superscripts $(\cdot)^T$ and $(\cdot)^H$ stand for the transpose and Hermitian transpose operations. $\text{diag}\{\cdot\}$ denotes a diagonal matrix. The notation $\otimes$ refers to Kronecker product.
The vectorization of a $M\times N$ matrix $\mathbf{X}$, denoted by $\text{vec}(\mathbf{X})$, is the $MN \times 1$ column vector $\mathbf{x}$ by stacking the columns of the matrix $\mathbf{X}$ on top of one another. The operator $\text{vec}^{-1}(\mathbf{x})$ is the inverse opration of the vectorization and $\text{vec}^{-1}(\mathbf{x})$ reshapes the $MN\times 1$ column vector $\mathbf{x}$ into a $M\times N$ matrix $\mathbf{X}$.
The notations $\mathbf{I}_M$ and $\mathbf{0}_{M\times N}$ represent $M \times M$ identity and $M\times N$ zero matrices, respectively.

\section{System Framework}\label{sec:system}

\begin{figure*}[t]
    \centering
    \includegraphics[width=0.8\textwidth]{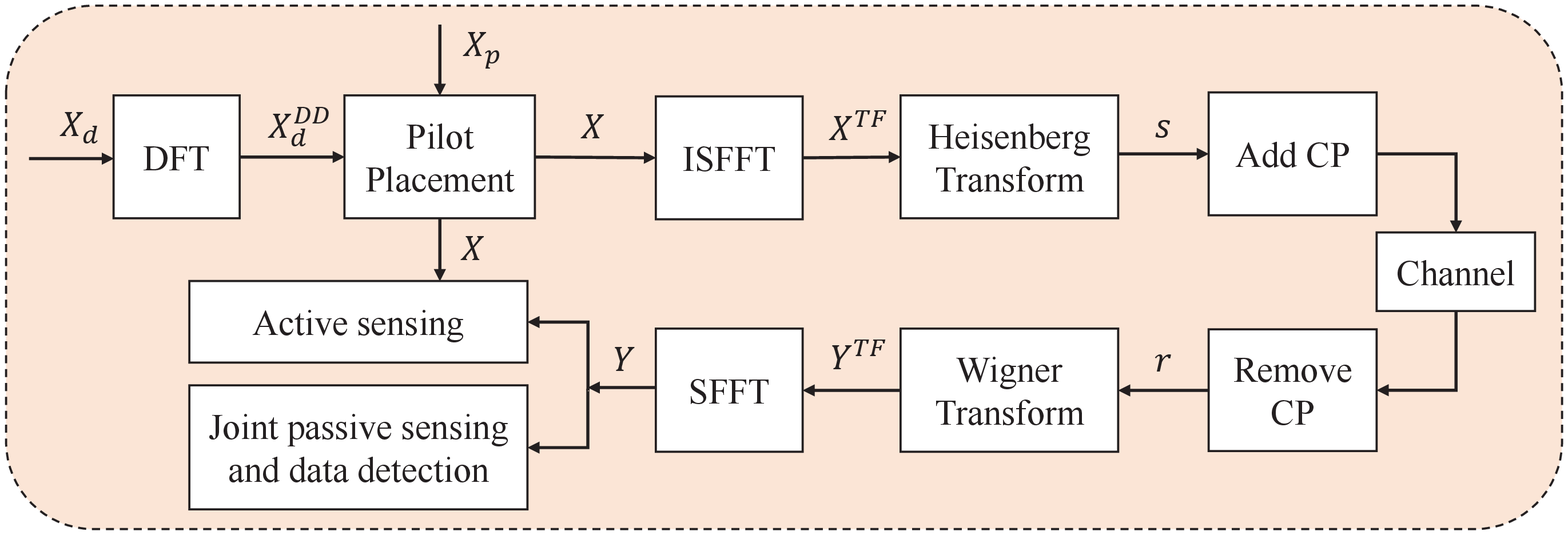}
    \caption{DFT-s-OTFS system model with superimposed pilot design for THz integrated sensing and communication.}
    \label{fig:dftsotfs-isac}
\end{figure*}

As shown in Fig.~\ref{fig:dftsotfs-isac}, we propose a sensing integrated DFT-spread OTFS system aided by superimposed pilots. At the transmitter, an ISAC waveform is generated and used for communication and sensing. For \textit{active sensing}, it bounces off the radar sensing targets and forms the back-scattered signal. The co-located receiver related to the transmitter captures the signal and estimates the target parameters with the full information of transmitted signal.
For \textit{passive sensing}, the transmitted signal propagates through a LoS path and several NLoS paths reflected by the sensing targets and conveys messages to the communication receiver, which performs joint passive sensing and data detection based on the knowledge of superimposed pilots.

\subsection{Transmitter Design and Processing}

First, the ISAC system maps the transmitted bit streams to a large amount of data frames, each of which contains $M \times N$ information symbols from a $Q$-ary quadrature amplitude modulation (QAM) alphabet $\mathbb{A}$, $\mathbf{X}_d\in \mathbb{C}^{M\times N}$. Here $M$ and $N$ stand for the number of delay and Doppler bins in the delay-Doppler domain. Equivalently, $M$ and $N$ denote the number of subcarriers and symbols in the time-frequency domain.
The data symbols are zero-mean independent and identically distributed (i.i.d.) with the signal power $\mathbb{E}\{|X_d[m, n]|^2\} = \sigma_d^2$, where $X_d[m, n]$ is the element of $\mathbf{X}_d$ at the $m$\textsuperscript{th} row and the $n$\textsuperscript{th} column, $m = 0, 1, \cdots, M - 1, n = 0, 1, \cdots, N - 1$.

\begin{figure*}
    \centering
    \subfigure[Design of pilot placement in delay-Doppler domain.]{\includegraphics[width=0.28\textwidth]{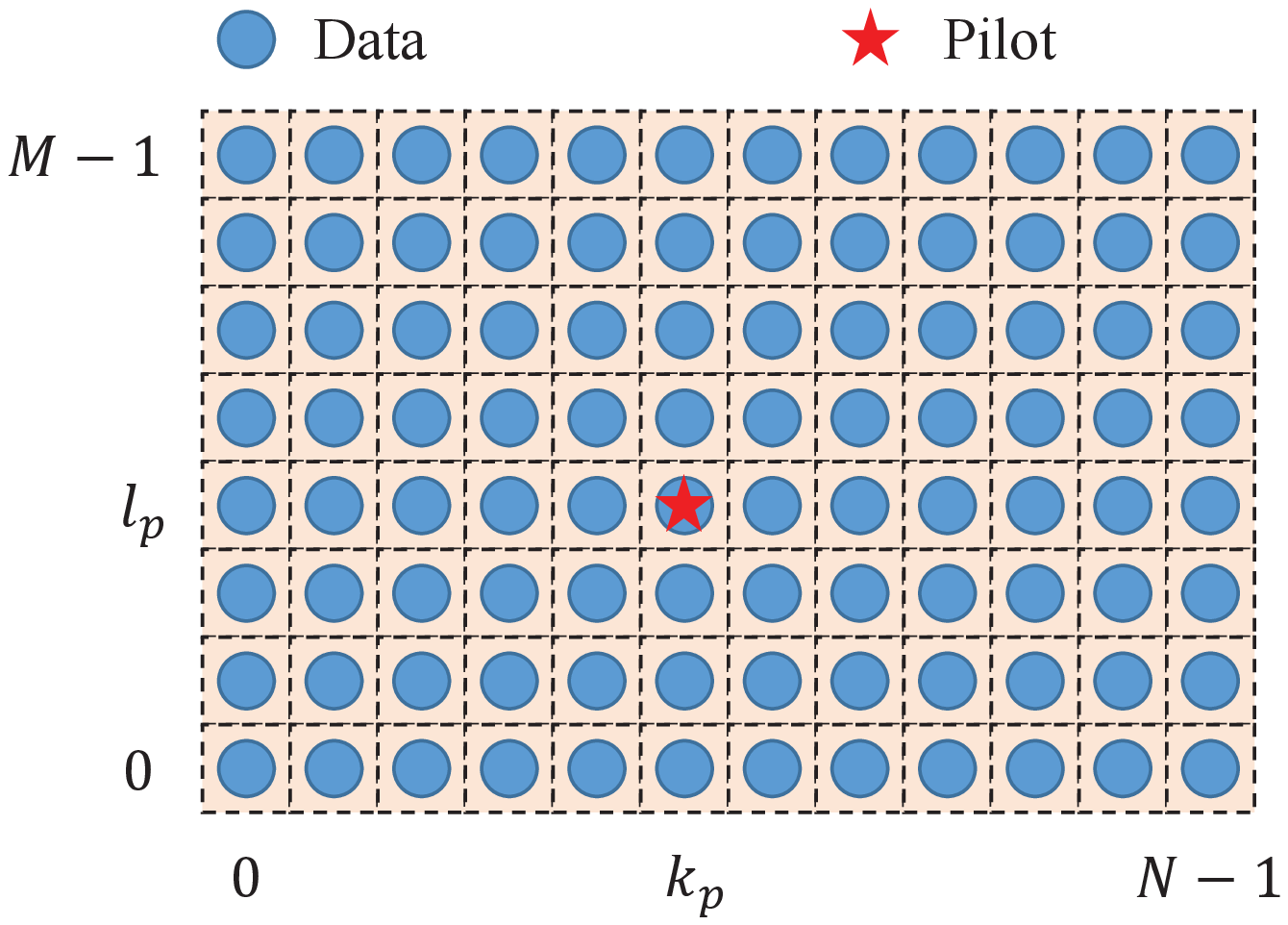}}
    \subfigure[Transmit signal in delay-Doppler domain.]{\includegraphics[width=0.32\textwidth]{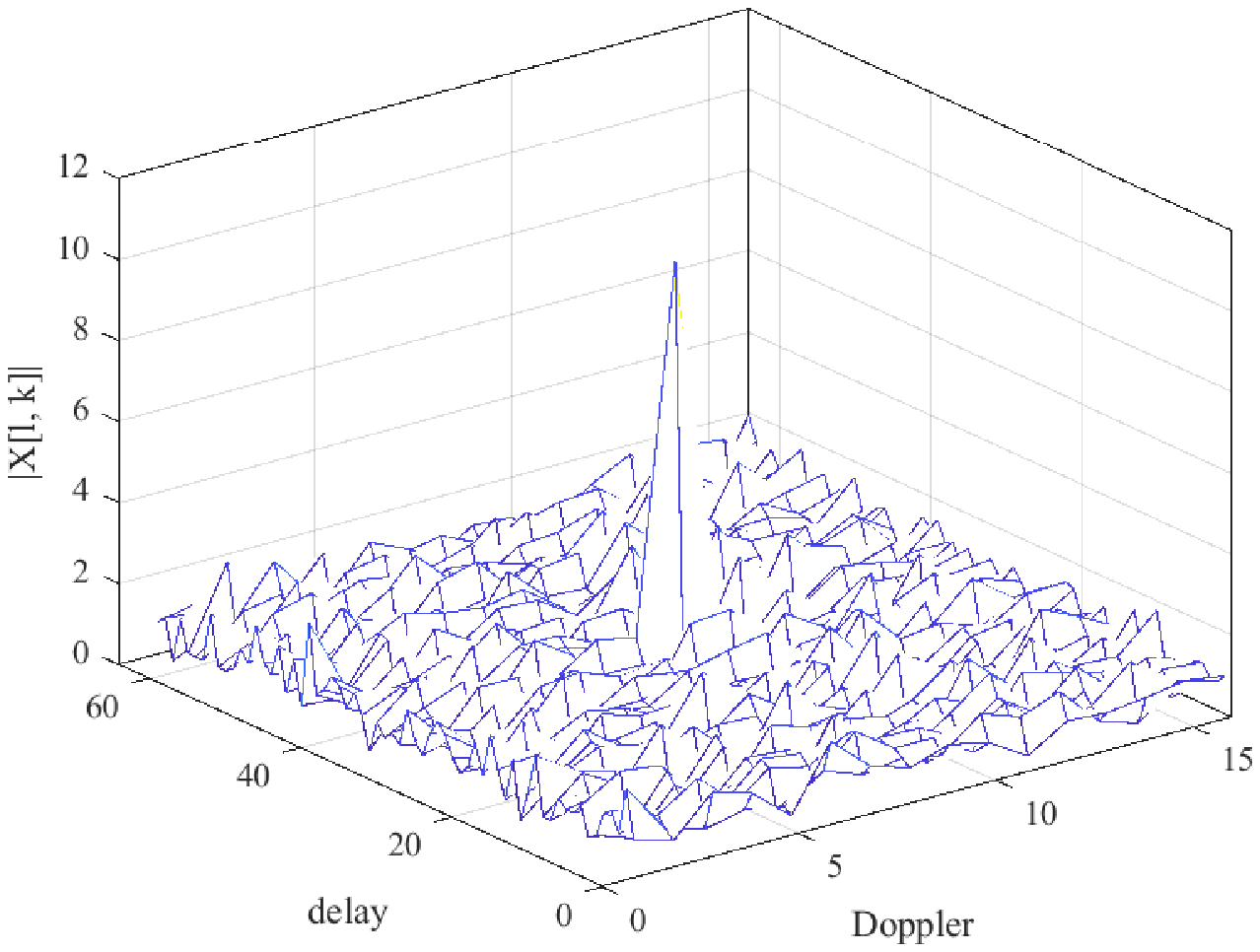}}
    \subfigure[Received signal in delay-Doppler domain with the 2-tap channel.]{\includegraphics[width=0.32\textwidth]{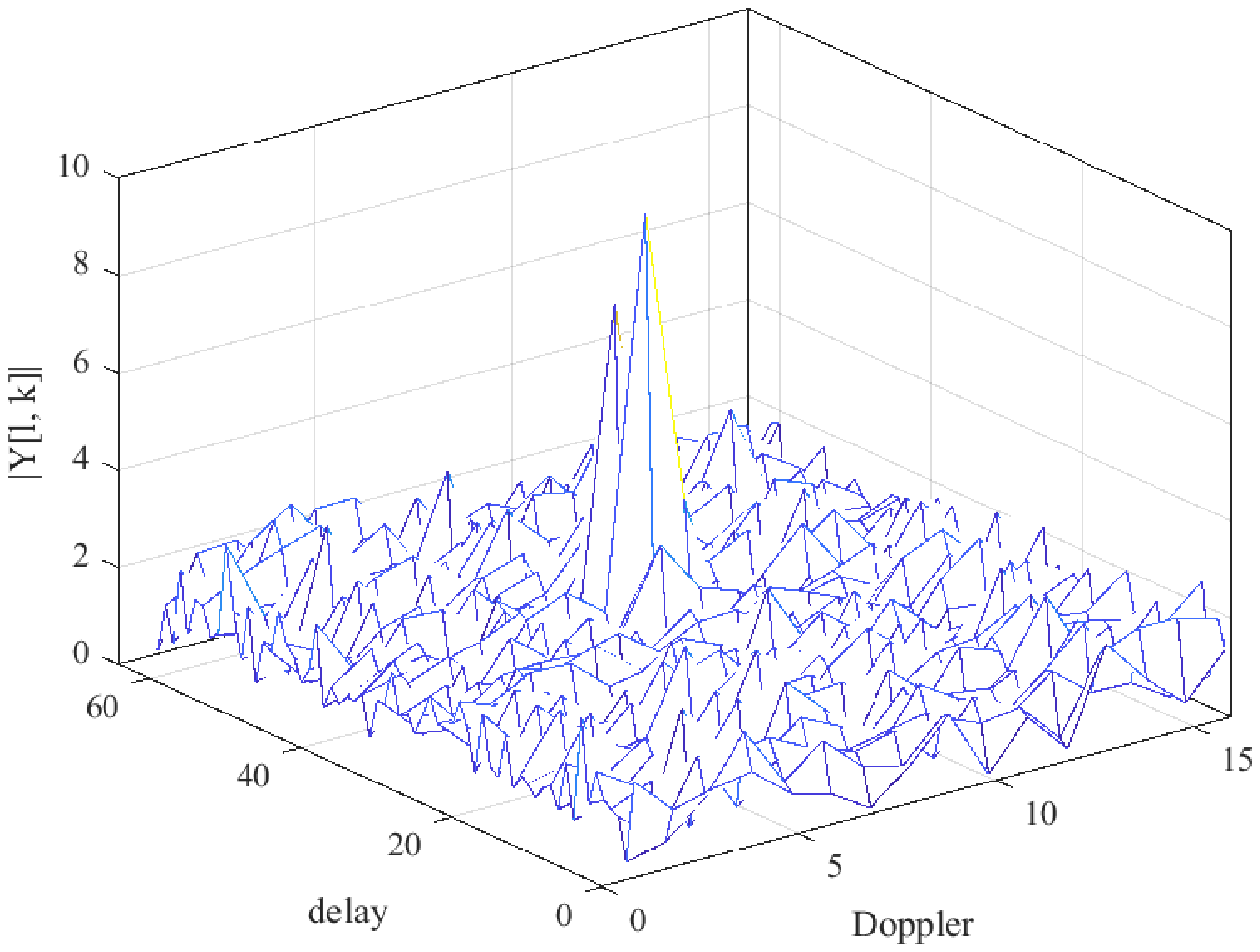}}
    \caption{Illustration of proposed frame structure with superimposed pilot and data symbols in DFT-s-OTFS systems.}
    \label{fig:superimposed_pilot}
\end{figure*}

Before mapping the information symbols onto the delay-Doppler plane, we perform a $N$-point DFT operation on the data symbols along the Doppler axis, as
\begin{equation}
    X_d^{\text{DD}}[l, k] = \frac{1}{\sqrt{N}} \sum_{n=0}^{N-1} X_d[l, n] e^{-j2\pi \frac{n k}{N}},
\end{equation}
for $l = 0, 1, \cdots, M - 1, k = 0, 1, \cdots, N - 1$. Then the pilot symbols $\mathbf{X}_p$ are superimposed onto the data symbols $\mathbf{X}_d^{\text{DD}}$ in the delay-Doppler domain, as
\begin{equation}
    \mathbf{X} = \mathbf{X}_d^{\text{DD}} + \mathbf{X}_p,
\end{equation}
where the pilot symbols are arranged as
\begin{equation}
    X_p[l, k] = \begin{cases}
    & \sqrt{MN \sigma_p^2}, \text{ } (l, k) = (l_p, k_p), \\
    & 0, \text{otherwise}.
    \end{cases}
\end{equation}
where $0 \leq l_p \leq M - 1, 0 \leq k_p \leq N - 1$ represent the pilot placement location in the delay-Doppler plane. The average pilot symbol power is defined as $\mathbb{E}\{|X_p[l, k]|^2\} = \sigma_p^2$.

As shown in Fig.~\ref{fig:superimposed_pilot}(a), the data symbols are arranged onto the whole delay-Doppler plane, while the non-zero pilot symbol is placed at only one delay-Doppler grid, i.e., $(l_p, k_p)$, which is superimposed onto the data symbol $X_d^{\text{DD}}[l_p, k_p]$. Zero-padding is performed for pilots at other delay-Doppler grids. Compared to the EP scheme that arranges a guard region around the EP with the insertion of zeros~\cite{raviteja2019pilot}, there is no dedicated grid for the pilot arrangement in the superimposed pilot scheme. Thus, the superimposed pilot design in this work reduces the pilot overhead and is able to improve the spectral efficiency of DFT-s-OTFS. As shown in Fig.~\ref{fig:superimposed_pilot}(b) and Fig.~\ref{fig:superimposed_pilot}(c), the pilot symbol has stronger power than data symbols and performs like a flag. Thus, the pilots in the received delay-Doppler domain signal can be used to estimate the channel parameters. Furthermore, we can develop an optimal allocation scheme between pilot and data. Since it is influenced by the data detection, the derivation is detailed in Sec.~\ref{sec:power_allocation}.

Next, the transmitter transforms the superimposed delay-Doppler domain symbols to the transmit signal $\mathbf{X}^{\text{TF}}$ in the time-frequency domain by applying the inverse symplectic finite Fourier transform (ISFFT)~\cite{raviteja2018otfs},
\begin{equation}
    X^{\text{TF}}[m, n]=\frac{1}{\sqrt{M N}} \sum_{k=0}^{N-1} \sum_{l=0}^{M-1} X[l, k] e^{j 2 \pi\left(\frac{n k}{N}-\frac{m l}{M}\right)},
\end{equation}
for $m = 0, 1, \cdots, M - 1, n = 0, 1, \cdots, N - 1$.
We denote the subcarrier spacing and the symbol duration of the time-frequency data frame as $\Delta f$ and $T$, where $T \Delta f = 1 $. In this case, each data frame occupies a frame duration of $T_s = N T$ and a bandwidth of $B = M \Delta f$.

Then, we use the Heisenberg transform to transform the 2D time-frequency domain symbols to the baseband time-domain transmitted signal, as
\begin{equation}\label{eq:baseband}
    s(t)= \frac{1}{\sqrt{M}}\sum_{n=0}^{N-1} \sum_{m=0}^{M-1} X^{\text{TF}}[m, n] g_{t x}(t-n T) e^{j 2 \pi m \Delta f(t-n T)},
\end{equation}
where $g_{tx}(t)$ stands for the transmit pulse shape that is limited to $[0, T]$. We consider to use practical rectangular transmit and receive pulses, which are compatible
with the OFDM modulation. Finally, one CP is inserted into the time-domain signal for each data frame, denoted by,
\begin{equation}
    s_{\text{CP}}(t)= \begin{cases}s(t), & 0 \leqslant t \leqslant T_s, \\ s(t+ T_s), & -T_{\text{cp}} \leqslant t < 0,\end{cases}
\end{equation}
where $T_\text{cp}$ denotes the CP duration. The transmit signal $s(t)$ can be expressed in matrix form as, $\mathbf{S} = \mathbf{F}_M^H \mathbf{X}^{\text{TF}} = \mathbf{F}_M^H ( \mathbf{F}_M \mathbf{X} \mathbf{F}_N^H) = \mathbf{X} \mathbf{F}_N^H$, with a sampling rate of $\frac{M}{T}$. Moreover, $\mathbf{F}_M\in \mathbb{C}^{M\times M}$ and $\mathbf{F}_N \in \mathbb{C}^{N \times N}$ refer to the normalized DFT matrices. Thus, the time-domain transmit vector $\mathbf{s}$ is given by
\begin{equation}
    \mathbf{s} = \text{vec}(\mathbf{S}) = \left(\mathbf{F}_N^H \otimes \mathbf{I}_M \right) \mathbf{x},
\end{equation}
where $\mathbf{x} = \text{vec}(\mathbf{X}) = \mathbf{x}_d^{\text{DD}} + \mathbf{x}_p$, $\mathbf{x}_d^{\text{DD}} $ and $\mathbf{x}_p$ are data and pilot vectors in the DD domain.

\subsection{Continuous-Delay-and-Doppler-Shift Channel Impulse Response}\label{sec:CDDS}

The DD domain channel response is characterized by sensing targets for a sensing channel or by transmission paths for a communication channel, which has a unified form for these two applications. We suppose that there exist $P$ multipath components, where $i$\textsuperscript{th} path is associated with complex path gain $\alpha_i$, delay $\tau_i$ and Doppler shift $\nu_i$. Here $\tau_i \in [0, \frac{1}{\Delta f})$, $\nu_i \in [-\frac{1}{2 T}, \frac{1}{2 T})$, and any two paths are resolvable in the delay-Doppler domain (i.e., $|\tau_i - \tau_j| \geq \frac{1}{M\Delta f}$ or $|\nu_i - \nu_j| \geq \frac{1}{NT}$). The impulse response of the wireless channel in the DD domain is thus given by
\begin{equation}
    h(\tau, \nu) = \sum_{i=1}^P \alpha_i \delta(\tau - \tau_i) \delta(\nu - \nu_i).
\end{equation}
For joint passive sensing and communication, the delay and Doppler shifts are calculated by $\tau_i = \frac{r_i}{c_0}$ and $\nu_i = \frac{f_c v_i}{c_0}$, where $r_i$ and $v_i$ refer to the distance and velocity along the $i$\textsuperscript{th} path. $f_c$ stands for the carrier frequency and $c_0$ represents the speed of light. For active sensing, the round-trip delay and Doppler shift lead to an extra multiplier of 2 in the above calculations.

When the path delay and Doppler shift are integer multiples of the delay and Doppler resolution, i.e., $\tau_i = \frac{l_i}{M\Delta f}$ and $\nu_i = \frac{k_i}{NT}$, the discrete baseband received signal vector $\mathbf{r}\in \mathbb{C}^{MN \times 1}$ is given by $\mathbf{r} = \mathbf{H} \mathbf{s} + \mathbf{w}$, where the channel $\mathbf{H} = \sum_{i = 1}^P \alpha_i \mathbf{\Delta}^{k_i} \mathbf{\Pi}_{MN}^{l_i} $~\cite{raviteja2019pulse}. The matrix $\mathbf{\Pi}_{MN} \in \mathbb{C}^{MN \times MN}$ denotes the forward cyclic-shift (permutation) matrix and $\mathbf{\Delta} = \text{diag}\{e^{j2\pi \frac{0}{MN} }, e^{j2\pi \frac{1}{MN} },\\
\cdots, e^{j2\pi \frac{MN - 1}{MN} }\}$ characterizes the Doppler shift. $\mathbf{w} \in \mathbb{C}^{MN \times 1}$ stands for the additive white Gaussian noise (AWGN) with zero mean and variance $\sigma_w^2$. While the integer case of delay and Doppler shift is well investigated in the literature~\cite{liu2021otfs, tiwari2019otfs},
low-complexity channel estimation and data detection with fractional channel and a rectangular pulse are still challenging for DFT-s-OTFS systems.
Thus, we consider the path delay and Doppler shift to be continuous-valued and derive a continuous-delay-and-Doppler-shift (CDDS) channel matrix.

The baseband received signal vector $\mathbf{r}$ is the sampling signal of the time-domain continuous signal $r(t)$ at $t = m\frac{T}{M}$ for $m = 0, 1, \cdots, MN - 1$, where $r(t)$ is given by
\begin{equation}\label{eq:rt}
\begin{split}
        r(t) &= \sum_{i=1}^{P} \alpha_i e^{j2\pi \nu_i t} s_\text{CP}(t - \tau_i) + w(t) \\
        &= \sum_{i=1}^{P} \alpha_i e^{j2\pi \nu_i t} s([t - \tau_i]_{NT}) + w(t),
\end{split}
\end{equation}
where $[\cdot]_T$ denotes modulo $T$ operation, $w(t)$ represents the AWGN.

Let $S_{\tau_i}[l, k]$ be the sampling signal of $s([t-\tau_i]_{NT})$ at $t = kT + \frac{l}{M} T$ for $l = 0, 1, \cdots, M - 1, k = 0, 1, \cdots, N - 1$, we obtain
    \begin{equation}
    \begin{split}
        S_{\tau_i}[l, k] =& \frac{1}{\sqrt{M}}\sum_{n=0}^{N-1} \sum_{m=0}^{M-1} X^{\text{TF}}[m, n] g_{t x}([kT + \frac{l}{M} T-\tau_i]_{NT} \\
        &- n T ) e^{j 2 \pi m \Delta f\left(\left[kT + \frac{l}{M} T-\tau_i\right]_{NT}-n T\right)} \\
        =& \frac{1}{\sqrt{M}}\sum_{n=0}^{N-1} \sum_{m=0}^{M-1} X^{\text{TF}}[m, n] g_{t x}(([k + \frac{l}{M} - \frac{\tau_i}{T}]_{N} \\
        &-n) T ) e^{j 2 \pi m ([k + \frac{l}{M} - \frac{\tau_i}{T}]_{N}-n)}.
    \end{split}
    \end{equation}
When $l < l_i = \lceil \frac{\tau_i}{\frac{T}{M}} \rceil$ ($\lceil \cdot \rceil$ stands for the ceiling function), $S_{\tau_i}[l, k] = \frac{1}{\sqrt{M}} \sum_{m=0}^{M-1} X^{\text{TF}}[m, [k - 1]_N] e^{j2\pi m (\frac{l}{M} - \frac{\tau_i}{T})}$. When $l \geqslant l_i$, $S_{\tau_i}[l, k] = \frac{1}{\sqrt{M}} \sum_{m=0}^{M-1} X^{\text{TF}}[m, k] e^{j2\pi m (\frac{l}{M} - \frac{\tau_i}{T})}$. Thus, we can derive its vector form as,
\begin{equation}
    \mathbf{s}_{\tau_i} = \mathbf{\Pi}_{MN}^{l_i} \text{vec}\left(\mathbf{\Pi}^{-l_i}_M \mathbf{F}_M^H \mathbf{b}_{\tau_i} \mathbf{X}^{\text{TF}} \right)
\end{equation}
where $\mathbf{b}_{\tau_i} = \text{diag}\{e^{-j2\pi0\frac{\tau_i}{T}}, \cdots, e^{-j2\pi(M-1)\frac{\tau_i}{T}}\}$. Following the eigendecomposition $\mathbf{\Pi}_M^{-l_i} = \mathbf{F}_M^H \mathbf{\Lambda} \mathbf{F}_M$ with $\mathbf{\Lambda} = \text{diag}\{e^{j2\pi0\frac{l_i}{M}}, \cdots, e^{j2\pi(M-1)\frac{l_i}{M}}\}$, we can obtain the relation between $\mathbf{s}_{\tau_i}$ and $\mathbf{s}$ as
\begin{equation}\label{eq:s_tau}
\begin{split}
   \mathbf{s}_{\tau_i} &= \mathbf{\Pi}_{MN}^{l_i} \text{vec}\left(\mathbf{F}_M^H \mathbf{\Lambda} \mathbf{F}_M \mathbf{F}_M^H \mathbf{b}_{\tau_i} \mathbf{F}_M \mathbf{S} \right) \\
   &= \mathbf{\Pi}_{MN}^{l_i} \text{vec}\left(\mathbf{F}_M^H \mathbf{B}_{\tau_i} \mathbf{F}_M \mathbf{S} \right) \\
   &= \mathbf{\Pi}_{MN}^{l_i} \left(\mathbf{I}_N \otimes \left(\mathbf{F}_M^H \mathbf{B}_{\tau_i} \mathbf{F}_M\right)  \right) \mathbf{s}, 
\end{split}
\end{equation}
where $\mathbf{B}_{\tau_i} = \text{diag}\{b^0, b^1, \cdots, b^{M-1}\}$ with $b = e^{j 2\pi \left(\frac{l_i}{M} - \frac{\tau_i}{T}\right) }$. Meanwhile, by sampling at $t = m\frac{T}{M}$ for $m = 0, 1, \cdots, MN -1$, we obtain the received vector $\mathbf{r}$, which is given by,
\begin{equation}
    \mathbf{r} = \sum_{i=1}^P \alpha_i \mathbf{\Delta}^{(\nu_i)} \mathbf{s}_{\tau_i} + \mathbf{w},
\end{equation}
where $\mathbf{\Delta}^{(\nu_i)} = \text{diag}\{\delta^0, \delta^1, \cdots, \delta^{MN-1}\}$ with $\delta = e^{j2\pi \nu_p \frac{T}{M}}$. Finally, we derive the baseband time-domain input-output relation for the CDDS channel as,
\begin{equation}
\begin{split}
\mathbf{r} &= \mathbf{H} \mathbf{s} + \mathbf{w} \\
        &= \sum_{i=1}^P \alpha_i \mathbf{\Theta}_i \mathbf{s} + \mathbf{w},
\end{split}
\end{equation}
where $\mathbf{\Theta}_i \overset{\triangle}{=}\mathbf{\Delta}^{(\nu_i)} \mathbf{\Pi}_{MN}^{l_i} \left(\mathbf{I}_N \otimes \left(\mathbf{F}_M^H \mathbf{B}_{\tau_i} \mathbf{F}_M\right)  \right)$.


\subsection{Receiver Design and Processing}

At the receiver, the received signal is transformed to the time-frequency domain signal by applying the Wigner transform, which is equivalent to DFT when using rectangular pulse, expressed in matrix form as, $\mathbf{Y}^{\text{TF}} = \mathbf{F}_M \mathbf{R}$ with $\mathbf{R} = \text{vec}^{-1}(\mathbf{r})\in \mathbb{C}^{M\times N}$. Then the symplectic finite Fourier transform (SFFT) transforms $\mathbf{Y}^{\text{TF}}$ back to the delay-Doppler domain, given by
\begin{equation}
\begin{split}
        \mathbf{Y} &= \mathbf{F}_M^H \mathbf{Y}^{\text{TF}} \mathbf{F}_N \\
        &= \mathbf{R} \mathbf{F}_N.
\end{split}
\end{equation}
The received vector in the delay-Doppler domain $\mathbf{y} = \text{vec}(\mathbf{Y})$ is derived as,
\begin{equation}\label{eq:dd_io}
\begin{split}
        \mathbf{y} &= \mathbf{A} \mathbf{x} + \mathbf{\tilde{w}} \\
        &= \mathbf{A} \mathbf{x}_d^{\text{DD}} + \mathbf{A} \mathbf{x}_p + \mathbf{\tilde{w}},
\end{split}
    \end{equation}
where the noise vector $\mathbf{\tilde{w}} = \left(\mathbf{F}_N \otimes \mathbf{I}_M\right) \mathbf{w}$, and the effective delay-Doppler domain channel matrix $\mathbf{A} \in \mathbb{C}^{MN\times MN}$ is expressed as
\begin{equation}
    \begin{split}
        \mathbf{A} &= \left(\mathbf{F}_N \otimes \mathbf{I}_M\right) \mathbf{H} \left(\mathbf{F}_N^H \otimes \mathbf{I}_M\right) \\
        &= \sum_{i=1}^P \alpha_i \mathbf{\Gamma}_i(\tau_i, \nu_i),
    \end{split}
\end{equation}
where $\mathbf{\Gamma}_i(\tau_i, \nu_i) \overset{\triangle}{=} \left(\mathbf{F}_N \otimes \mathbf{I}_M\right)  \mathbf{\Theta}_i \left(\mathbf{F}_N^H \otimes \mathbf{I}_M\right) $. The channel matrix $\mathbf{A}$ depends on the channel parameters $(\bm{\alpha}, \bm{\tau}, \bm{\nu})$, where $\bm{\alpha} = [\alpha_1, \alpha_2, \cdots, \alpha_P]^T \in \mathbb{C}^{P\times 1}$, $\bm{\tau} = [\tau_1, \tau_2, \cdots, \tau_P]^T \in \mathbb{R}_{+}^{P\times 1}$, $\bm{\nu} = [\nu_1, \nu_2, \cdots, \nu_P]^T \in \mathbb{R}^{P\times 1}$.

After the receiver processing, we are able to conduct two modes for the ISAC applications, i.e., active sensing, joint passive sensing and data detection. The tasks of these two modes are described as follows:
\begin{itemize}
    \item \textit{Active sensing:} The objective is to estimate the channel delay $\bm{\tau}$ and the Doppler shifts $\bm{\nu}$, given the transmit vector $\mathbf{x}$ and the received vector $\mathbf{y}$.
    \item \textit{Joint passive sensing and data detection:} The objective is to estimate the channel parameters $(\bm{\alpha}, \bm{\tau}, \bm{\nu})$ and recover $\mathbf{X}_d$ (or $\mathbf{x}_d^{\text{DD}}$), given the pilot vector $\mathbf{x}_p$ and the received vector $\mathbf{y}$.
\end{itemize}

\section{Active Sensing Parameter Estimation}\label{sec:active_sensing}

In this section, we propose the sensing parameter estimation algorithm.
For active sensing receivers, e.g., a radar sensing receiver, the transmit vector $\mathbf{x}$ is known and the sensing parameters can be estimated with full knowledge of the transmit signal, including the data vector $\mathbf{x}_d^{\text{DD}}$ and pilot vector $\mathbf{x}_p$~\cite{gaudio2020otfs}.
Specifically, we develop a two-phase search method based on the maximum likelihood estimator (MLE), where in the first phase we perform on-grid search with coarse estimation and in the second phase we conduct off-grid search to refine the estimation result.

\subsection{Maximum Likelihood Estimator}

Based on the CDDS channel model in Sec.~\ref{sec:CDDS} with $P$ targets, we aim to obtain the estimation results for the set of 3 unknown vectors with $3P$ parameters, $(\bm{\alpha}, \bm{\tau}, \bm{\nu})$, which are used to calculate the target range and velocity.
By minimizing the log-likelihood function, the maximum likelihood estimator of these parameters is given by
\begin{equation}
    (\hat{\bm{\alpha}}, \hat{\bm{\tau}}, \hat{\bm{\nu}}) = \arg\min_{(\bm{\alpha}, \bm{\tau}, \bm{\nu})} \left\|\sum_{i=1}^P \alpha_i \mathbf{\Gamma}_i \mathbf{x} - \mathbf{y}\right\|^2
\end{equation}
A direct search for these parameters in a $3P$-dimensional continuous domain $\mathbb{C}^{P}\times \mathbb{R}_{+}^{P} \times \mathbb{R}^{P} $ requires high complexity and becomes intractable in multi-target sensing. Thus, we transform the MLE into $P$ estimators, each of which performs one search in a 2-dimensional domain. Meanwhile, we propose the following two-phase method with super-resolution estimation accuracy and low complexity.

\subsection{Two-Phase Sensing Parameter Estimation Method}

For the $i$\textsuperscript{th} target, $\alpha_i \mathbf{\Gamma}_i \mathbf{x}$ and $\sum_{j\neq i} \alpha_j \mathbf{\Gamma}_j \mathbf{x}$ denote the useful signal and interference signal. Thus, we need to perform interference cancellation mechanism, i.e., when estimating the parameters of $i$\textsuperscript{th} target, eliminating the interference signal from the $(i - 1)$ targets that have been estimated. To be specific, for the first target with parameters $(\alpha_1, \tau_1, \nu_1)$, the estimator is given by
\begin{equation}
    (\hat{\alpha}_1, \hat{\tau}_1, \hat{\nu}_1) = \arg\min_{(\alpha_1, \tau_1, \nu_1)} \left\|\alpha_1 \mathbf{\Gamma}_1 \mathbf{x} - \mathbf{y}\right\|^2.
\end{equation}
Next, this minimization problem boils down to the maximization problem,
\begin{equation}
    (\hat{\tau}_1, \hat{\nu}_1) = \arg\max_{( \tau_1, \nu_1)} \left|(\mathbf{\Gamma}_1 \mathbf{x} )^H\mathbf{y}\right|^2,
\end{equation}
and the channel coefficient $\alpha_1$ is calculated by
$
    \hat{\alpha}_1 = \frac{\mathbf{x}^H \mathbf{\Gamma}_1^H(\hat{\tau_1}, \hat{\nu_1}) \mathbf{\Gamma}_1(\hat{\tau_1}, \hat{\nu_1}) \mathbf{x}}{(\mathbf{\Gamma}_1(\hat{\tau_1}, \hat{\nu_1}) \mathbf{x})^H \mathbf{y}}
$.
For the remaining $(P - 1)$ targets, e.g., the $i$\textsuperscript{th} target, $i = 2, 3, \cdots, P$, the interference cancellation is performed by subtracting the interference signal $\sum_{j = 1}^{i-1} \hat{\alpha}_j \mathbf{\Gamma}_j(\hat{\tau}_j, \hat{\nu}_j) \mathbf{x} $ from the received vector $\mathbf{y}$. In this case, the estimator for the $i$\textsuperscript{th} target becomes
\begin{equation}\label{eq:2d_estimator}
    (\hat{\tau}_i, \hat{\nu}_i) = \arg\max_{( \tau_i, \nu_i)} \left|(\mathbf{\Gamma}_i \mathbf{x} )^H \left(\mathbf{y} - \sum_{j = 1}^{i-1} \hat{\alpha}_j \mathbf{\Gamma}_j(\hat{\tau}_j, \hat{\nu}_j) \mathbf{x} \right)\right|^2,
\end{equation}
and the estimated channel coefficient $\alpha_i$ is derived as
\begin{equation}\label{eq:alpha}
    \hat{\alpha}_i = \frac{\mathbf{x}^H \mathbf{\Gamma}_i^H(\hat{\tau_i}, \hat{\nu_i}) \mathbf{\Gamma}_i(\hat{\tau_i}, \hat{\nu_i}) \mathbf{x}}{(\mathbf{\Gamma}_i(\hat{\tau_1}, \hat{\nu_1}) \mathbf{x})^H \left(\mathbf{y}- \sum_{j = 1}^{i-1} \hat{\alpha}_j \mathbf{\Gamma}_j(\hat{\tau}_j, \hat{\nu}_j) \mathbf{x}\right)}.
\end{equation}
To interpret, the 2D search problem in \eqref{eq:2d_estimator} is to find $(\tau_i, \nu_i)$ in the region $[0, \frac{1}{\Delta f}) \times [-\frac{1}{2 T}, \frac{1}{2 T})$ at which the objective function of \eqref{eq:2d_estimator} has a maximum. In order to reduce the searching complexity, we consider to use interval-reducing methods and narrow the region of uncertainty that contains the true values of $\tau_i$ and $\nu_i$. As an overview, in the first phase, initial bracketing of the maximum is performed by searching using a uniform grid. In the second phase, we evaluate the points in the established region in Phase I. We describe the proposed two-phase estimator in the following.

\subsubsection{Phase I}

To reduce the region of uncertainty, we propose on-grid parameter search on the discretized grid $\Lambda = \left\{\left(\frac{l}{M \Delta f}, \frac{k}{N T}\right), l=0, \cdots, M-1, k= - \frac{N}{2}, \cdots, \frac{N}{2}-1\right\}$ in the delay-Doppler plane. When the delay and Doppler are integer multiples of the delay and Doppler resolution, $\tau_i = \frac{l_i}{M \Delta f}, \nu_i = \frac{k_i}{NT}$, the received symbols in the delay-Doppler domain can be approximately expressed as the 2D circular shift of the transmitted symbols~\cite{raviteja2018otfs}, as
\begin{equation}
    \begin{split}
        Y[l, k] \approx & \sum_{i=1}^P \alpha_i e^{j2\pi\left(\frac{l-l_i}{M}\right)\frac{k_i}{N}} \beta_i(l, k) X[[l - l_i]_M, [k - k_i]_N] \\
        &+ \tilde{W}[l, k],
    \end{split}
\end{equation}
where
$\beta_i(l, k)= \begin{cases}1 & l_{i} \leq l<M, \\ \frac{N-1}{N} e^{-j 2 \pi\left(\frac{\left[k-k_{i}\right]_N}{N}\right)} & 0 \leq l<l_{i}.\end{cases}$ and $\tilde{W}[l, k]$ represents the noise. Then we obtain the modified maximization estimator of \eqref{eq:2d_estimator} as,
\begin{equation}\label{eq:phase1}
    (\hat{l}_i, \hat{k}_i) = \arg\max_{(l_i, k_i) \in \Lambda } \left|\text{vec}\left(\mathbf{\Pi}_M^{l_i}\mathbf{X}\mathbf{\Pi}_{N}^{-k_i}\right)^H\mathbf{y}_{ni}\right|^2,
\end{equation}
where $\mathbf{y}_{ni} \overset{\triangle}{=} \mathbf{y} - \sum_{j = 1}^{i-1} \hat{\alpha}_j \mathbf{\Gamma}_j(\hat{\tau}_j, \hat{\nu}_j) \mathbf{x} $. We take the estimated delay $\hat{\tau}_i$ to lie between the points $\frac{\hat{l}_i - 1}{M\Delta f}$ and $\frac{\hat{l}_i + 1}{M\Delta f}$, and the estimated Doppler $\hat{\nu}_i$ to lie between $\frac{\hat{k}_i - 1}{N T}$ and $\frac{\hat{k}_i + 1}{N T}$. Thus, lower and upper limits for the region of uncertainty are established and the search region for Phase II becomes,
\begin{equation}
    \Lambda_i = \left\{(\tau, \nu), \frac{\hat{l}_i - 1}{M\Delta f} \leqslant\tau \leqslant \frac{\hat{l}_i + 1}{M\Delta f}, \frac{\hat{k}_i - 1}{N T} \leqslant \nu \leqslant \frac{\hat{k}_i + 1}{N T}  \right\}.
\end{equation}

\subsubsection{Phase II}

In this phase, we need to conduct off-grid search over the established region $\Lambda_i$ in Phase I, as
\begin{equation}\label{eq:phase2}
    \left(\hat{\tau}_i, \hat{\nu}_i\right) = \arg\max_{(\tau_i, \nu_i) \in \Lambda_i} \left|(\mathbf{\Gamma}_i \mathbf{x})^H \mathbf{y}_{ni}\right|^2.
\end{equation}
To solve this 2D maximization, we propose a 2D golden section method, which reduces the interval of uncertainty by the golden ratio for each step.
With the estimated delay and Doppler, the target range and velocity can be calculated as, $\hat{r}_i = \frac{\hat{\tau}_i c_0}{2}$, $\hat{v}_i = \frac{\hat{\nu}_i c_0}{2 f_c}$, respectively.

\begin{algorithm}[t]
\caption{Proposed Two-Phase Estimation Algorithm for Active Sensing}
\label{alg:TPE}
\KwIn{Received vector $\mathbf{y}$, transmit vector $\mathbf{x}$.}
\KwOut{Estimated channel parameters, $(\hat{\bm{\alpha}}, \hat{\bm{\tau}}, \hat{\bm{\nu}})$.}
\textit{Initialization:} $\mathbf{y}_{ni} = \mathbf{y}$, $\eta = \frac{\sqrt{5} - 1}{2}$\;
\For{$i = 1:P$}{
Solve \eqref{eq:phase1} and obtain $(\hat{l}_i, \hat{k}_i)$\;
$a_u = \frac{\hat{l}_i - 1}{M\Delta f}, a_l = \frac{\hat{l}_i + 1}{M\Delta f}, b_l = \frac{\hat{k}_i - 1}{N T}, b_u = \frac{\hat{k}_i + 1}{N T}$\;
\Repeat{Stopping criteria}{
$I_a = a_u - a_l, I_b = b_u - b_l$\;
$a_1 = a_l + (1 - \eta) I_a, a_2 = a_l + \eta I_a $\;
$b_1 = b_l + (1 - \eta) I_b, b_2 = b_l + \eta I_b$\;
$f_{11} = |(\mathbf{\Gamma}_i(a_1, b_1) \mathbf{x})^H \mathbf{y}_{ni}|^2, f_{12} = |(\mathbf{\Gamma}_i(a_1, b_2) \mathbf{x})^H \mathbf{y}_{ni}|^2, f_{21} = |(\mathbf{\Gamma}_i(a_2, b_1) \mathbf{x})^H \mathbf{y}_{ni}|^2, f_{22} = |(\mathbf{\Gamma}_i(a_2, b_2) \mathbf{x})^H \mathbf{y}_{ni}|^2$\;
\textbf{switch} $\max\{f_{11}, f_{12}, f_{21}, f_{22}\}$ \textbf{do}\\
\quad \textbf{case} $f_{11}$, $a_u = a_2, b_u = b_2$\;
\quad \textbf{case} $f_{12}$, $a_u = a_2, b_l = b_1$\;
\quad \textbf{case} $f_{21}$, $a_l = a_1, b_u = b_2$\;
\quad \textbf{case} $f_{22}$, $a_l = a_1, b_l = b_1$\;
\textbf{end}\\
}
$\hat{\tau}_i = \frac{a_l + a_u}{2}, \hat{\nu}_i = \frac{b_l + b_u}{2}$\;
Calculate $\hat{\alpha}_i$ according to \eqref{eq:alpha}\;
Update $\mathbf{y}_{ni} \leftarrow \mathbf{y}_{ni} - \hat{\alpha}_{i} \mathbf{\Gamma}_{i}(\hat{\tau}_{i}, \hat{\nu}_{i}) \mathbf{x}$\;
}
\end{algorithm}

\begin{figure}
    \centering
    \includegraphics[width=0.45\textwidth]{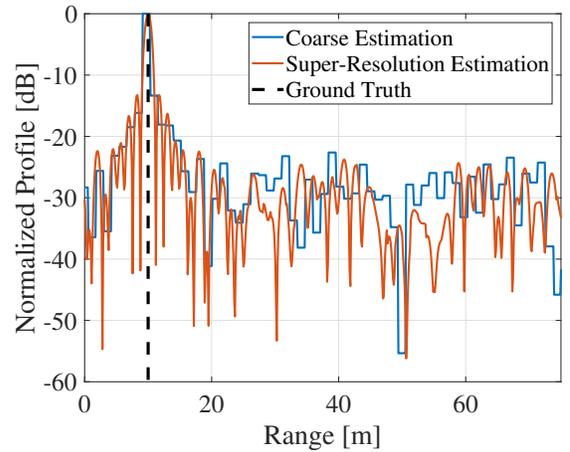}
    \caption{Normalized range profile of on-grid search in Phase I and off-grid search in Phase II with a sensing target at $r$ = 10 m.}
    \label{fig:range_profile}
\end{figure}

The two-phase method is summarized in Algorithm~\ref{alg:TPE}. For multi-target estimation, we eliminate the interference signal from the previous $(i-1)$ targets when estimating the parameters of $i$\textsuperscript{th} target. The estimation of each target includes two phases. At the first phase, the integer parts of delay and Doppler parameters are estimated by solving~\eqref{eq:phase1}, which reduces the search region of the second phase. At the second phase, the fractional delay and Doppler parameters are estimated by solving~\eqref{eq:phase2} with a 2D golden section search method and the channel coefficient parameters are calculated according to \eqref{eq:alpha}.

We provide an example to illustrate the proposed two-phase sensing parameter estimation algorithm. We consider a target with the range of 10 meters. The radar range profiles using on-grid search in Phase I and off-grid search in Phase II are plotted in Fig.~\ref{fig:range_profile}. 
The mathematical expressions to compute normalized profile are given by the objective functions of \eqref{eq:phase1} for coarse estimation and \eqref{eq:phase2} for super-resolution estimation with normalization.
We demonstrate that the profile of coarse estimation in Phase I has a flat ``peak" around the ground truth, while the curve of super-resolution estimation in Phase II has a steep peak at the ground truth. Therefore, we can first use coarse estimation in Phase I to reduce the search region with low computational complexity, and then apply super-resolution estimation in Phase II to obtain significantly improved results.

\subsection{Complexity Analysis}\label{sec:complexity}

We here analyze the complexity of the proposed two-phase estimation method. In Phase I, we need to compute $\text{vec}\left(\mathbf{\Pi}_M^{l_i}\mathbf{X}\mathbf{\Pi}_{N}^{-k_i}\right)$, which is performed by a 2D circular shift operation with low complexity. In Phase II, the key step is the computation of $\mathbf{\Gamma}_i \mathbf{x}$. Direct multiplication of the matrix and the vector is not computationally efficient, which has complexity of $\mathcal{O}(M^2 N^2)$. Instead, we can calculate $\mathbf{\Gamma}_i \mathbf{x}$ as,
\begin{equation}
    \text{vec}\left(\text{vec}^{-1}\left(\mathbf{\Delta}^{(\nu_i)} \mathbf{\Pi}_{MN}^{l_i} \text{vec}\left(\mathbf{F}_M^H \mathbf{B}_{\tau_i} \mathbf{F}_M \mathbf{X}\mathbf{F}_N^H \right) \right) \mathbf{F}_N\right).
\end{equation}
The above calculation only requires the multiplication operation with diagonal matrices $\mathbf{\Delta}^{(\nu_i)}, \mathbf{B}_{\tau_i}$, cyclic-shift operation and DFT/IDFT operation. The multiplication operation with a diagonal matrix uses $MN$ complex multiplications. The $N$-point DFT/IDFT operation on an $M\times N$ matrix has implementation complexity of $\mathcal{O}\left(MN\log(N)\right)$ by using the fast Fourier transform (FFT) algorithm, while the $M$-point DFT/IDFT has complexity of $\mathcal{O}(MN \log(M))$. Therefore, the overall computation complexity of the two-phase estimation method is $\mathcal{O}\left(MN\log(MN)\right)$.

In an existing sensing algorithm for OTFS~\cite{gaudio2020otfs}, a direct off-grid search in the whole continuous delay-Doppler domain requires high computational complexity. Our method performs on-grid estimation at the first phase by exploiting the input-output relation of 2D circular shift in the delay-Doppler domain, which significantly reduces the search region with low complexity. We conduct the off-grid search at the second phase by using the FFT algorithm, which can improve the computational efficiency compared with the off-grid method in~\cite{gaudio2020otfs} that directly calculates the multiplication of the $MN \times MN$ channel matrix and the $MN \times 1$ vector.

\section{Joint Passive Sensing and Data Detection}\label{sec:passive_sensing}

In this section, we focus on the joint passive sensing and data detection, where the communication receiver also serves as the sensing receiver. We aim at estimating the channel parameters based on the received signal and the information of transmit pilot vector, which are used for signal recovery and passive sensing. First, we employ the sensing parameter estimation method in Sec.~\ref{sec:active_sensing} aided by the pilot signal, to obtain the coarse estimation results of channel parameters. Then, we propose an iterative channel estimation and data detection to refine the estimation accuracy and detection performance. 

\subsection{Coarse Channel Estimation using Superimposed Pilots}

Since the pilot signal is known and the data signal is unknown at the communication receiver, we use superimposed pilots to estimate the channel parameters, by treating the received signal induced by data symbols as interference. The coarse estimator is formulated as
\begin{equation}\label{eq:coarse_ce}
    (\hat{\bm{\alpha}}^{(0)}, \hat{\bm{\tau}}^{(0)}, \hat{\bm{\nu}}^{(0)}) = \arg\min_{(\bm{\alpha}, \bm{\tau}, \bm{\nu})} \left\|\sum_{i=1}^P \alpha_i \mathbf{\Gamma}_i \mathbf{x}_p - \mathbf{y}\right\|^2,
\end{equation}
which can be solved by using Algorithm~\ref{alg:TPE}. Since only superimposed pilots are used for channel estimation, the estimation results are coarse and the estimation accuracy is worse than that by using full information of data and pilot symbols. Nevertheless, the estimated channel parameters $(\hat{\bm{\alpha}}^{(0)}, \hat{\bm{\tau}}^{(0)}, \hat{\bm{\nu}}^{(0)})$ can still be used to equalize the received data signal and obtain approximate recovered data symbols.

\begin{figure}
    \centering
    \includegraphics[width=0.45\textwidth]{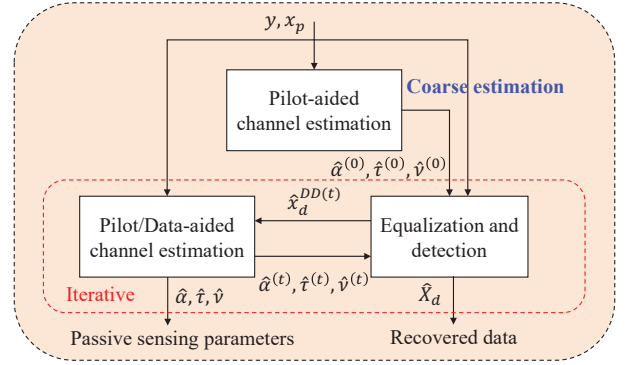}
    \caption{Block diagram for the joint passive sensing and data detection.}
    \label{fig:iterative_ce_dd}
\end{figure}

\subsection{Iterative Channel Estimation and Data Detection}

 With the detected data and superimposed pilots, more accurate channel estimation can be realized to further refine the performance of data detection. As shown in Fig.~\ref{fig:iterative_ce_dd}, we develop an iterative channel estimation and data detection method. At each iteration, we first equalize the received signal, detect the data symbols $\hat{\mathbf{X}}_d^{(t)}$ and obtain the estimated delay-Doppler domain signal $\hat{\mathbf{X}}_d^{DD(t)}$ based on the estimation results of last iteration $(\hat{\bm{\alpha}}^{(t-1)}, \hat{\bm{\tau}}^{(t-1)}, \hat{\bm{\nu}}^{(t-1)})$. Next, we design a pilot- and data-aided channel estimator with the estimated transmit data symbols,
\begin{equation}\label{eq:iterative_ce}
    (\hat{\bm{\alpha}}^{(t)}, \hat{\bm{\tau}}^{(t)}, \hat{\bm{\nu}}^{(t)}) = \arg\min_{(\bm{\alpha}, \bm{\tau}, \bm{\nu})} \left\|\sum_{i=1}^P \alpha_i \mathbf{\Gamma}_i \hat{\mathbf{x}}^{(t)}- \mathbf{y}\right\|^2,
\end{equation}
where $\hat{\mathbf{x}}^{(t)} = \mathbf{x}_p + \hat{\mathbf{x}}_d^{DD(t)}$, $\hat{\mathbf{x}}_d^{DD(t)} = \text{vec}\left(\hat{\mathbf{X}}_d^{DD(t)}\right)$ denotes the estimated DD domain data vector.

Recalling \eqref{eq:dd_io}, the channel equalization is to recover $\mathbf{x}$ from the received vector $\mathbf{y}$ by employing the channel matrix $\mathbf{A}$ (or channel parameters $\bm{\alpha}, \bm{\tau}, \bm{\nu}$). The channel equalization is developed by solving the $l_2$-regularized least squares program (LSP) problem as
\begin{equation}\label{eq:dd_ls}
    \min_{\mathbf{x}} \|\mathbf{A}\mathbf{x} - \mathbf{y}\|_2^2 + \lambda \|\mathbf{x}\|^2,
\end{equation}
where $\lambda$ equals to the inverse of the signal-to-noise ratio (SNR). Alternatively, a time-domain equalizer can be developed and expressed as a $l_2$-regularized LSP problem,
\begin{equation}\label{eq:t_ls}
    \min_{\mathbf{s}} \|\mathbf{H}\mathbf{s} - \mathbf{r}\|_2^2 + \lambda \|\mathbf{s}\|^2.
\end{equation}
Since the time-domain transmit signal $\mathbf{S}$ and the data signal $\mathbf{X}_d$ have the relation $\mathbf{S} = \mathbf{X}_d + \mathbf{X}_p \mathbf{F}_N^H$, the performance using the formulation with respect to $\mathbf{S}$ is indeed identical to that with respect to $\mathbf{X}_d$.
The minimum mean square error (MMSE) equalization methods can be employed to solve these two LSPs and have the analytical solutions, given by
\begin{equation}
    \hat{\mathbf{x}}_{\text{MMSE}} = \left(\mathbf{A}^H\mathbf{A} + \lambda \mathbf{I}\right)^{-1} \mathbf{A}^H \mathbf{y},
\end{equation}
\begin{equation}
    \hat{\mathbf{s}}_{\text{MMSE}} = \left(\mathbf{H}^H\mathbf{H} + \lambda \mathbf{I}\right)^{-1} \mathbf{H}^H \mathbf{r}.
\end{equation}
When the solutions to the above LSPs are computed by direct methods of matrix inversion, the computation complexity of $\mathcal{O}((MN)^3)$ is required. These analytical solutions are computationally inefficient, since the channel matrices $\mathbf{A}$ and $\mathbf{H}$ have a large size of $MN \times MN$. When the path delay and Doppler shift are integer multiples of the delay and Doppler resolution, the channel matrices become block-circulant. In this case, the matrix-decomposition techniques are employed to develop a MMSE receiver for OTFS with the implementation complexity of $\mathcal{O}(MN \log(MN))$. Nevertheless, for CDDS channels with fractional delay and Doppler, the block-circulant property of the channel matrices is not satisfied and the matrix-decomposition methods cannot be applied. Instead, several detection algorithms based on message passing (MP) are proposed in~\cite{raviteja2018otfs,gaudio2020otfs,liu2021otfs} to reduce the complexity of OTFS data detection by calculating the posterior probability and directly making decisions on transmitted symbols. Since the information symbols of DFT-s-OTFS are not mapped on the delay-Doppler domain, the MP algorithms can not exploit the sparsity of delay-Doppler domain channel for DFT-s-OTFS.

\begin{algorithm}[t]
\caption{The conjugate gradient method for the channel equalizer of DFT-s-OTFS}
\label{alg:cg}
\KwIn{Time-domain received vector $\mathbf{r}$, channel parameters $\bm{\alpha}, \bm{\tau}, \bm{\nu}$, inverse of SNR $\lambda$.}
\KwOut{Estimated baseband time-domain transmit vector $\hat{\mathbf{s}}$.}
\textit{Initialization:} $\mathbf{s}_0 = \mathbf{0}_{MN\times 1}$, $\mathbf{r}_0 = \mathbf{r} - \mathbf{H} \mathbf{s}_0$, $\mathbf{x}_0 = \mathbf{p}_0 = \mathbf{H}^H \mathbf{r}_0$, $\gamma_0 = \|\mathbf{x}_0\|^2 $, $t = 0$\;
\Repeat{Stopping criteria}{
$\mathbf{q}_t = \mathbf{H} \mathbf{p}_t, \beta_t = \frac{\gamma_t}{\|\mathbf{q}_t\|^2 + \lambda \|\mathbf{p}_t\|^2} $\;
$\mathbf{s}_{t+1} = \mathbf{s}_{t} + \beta_t \mathbf{p}_t, \mathbf{r}_{t+1} = \mathbf{r}_t - \beta_t \mathbf{q}_t$\;
$\mathbf{x}_{t+1} = \mathbf{H}^H \mathbf{r}_{t+1} - \lambda \mathbf{s}_{t+1}$\;
$\gamma_{t+1} = \|\mathbf{x}_{t+1}\|^2$\;
$\mathbf{p}_{t+1} = \mathbf{x}_{t+1} + \frac{\gamma_{t+1}}{\gamma_{t}} \mathbf{p}_t$\;
$t = t + 1$\;
}
\Return{$\hat{\mathbf{s}} = \mathbf{s}_t$.}\;
\end{algorithm}

Large-scale $l_2$-regularized least squares problem in \eqref{eq:dd_ls} and \eqref{eq:t_ls} has been well investigated in mathematics and data analysis~\cite{saad2003iterative}. When the direct methods do not work well for the channel matrix with large size, iterative methods can be utilized to solve the linear system of equations $\left(\mathbf{H}^H\mathbf{H} + \lambda \mathbf{I}\right)\mathbf{s} = \mathbf{H}^H \mathbf{r}$. Iterative methods are computationally efficient, especially when fast algorithms can be employed for the matrix-vector multiplications $\mathbf{H}\mathbf{r}$ and $\mathbf{H}^H \mathbf{r}$. At the proposed DFT-s-OTFS receiver, the channel matrix has a special form, i.e., partial Fourier and cyclic-shift matrices. The matrix $\mathbf{H}$ is used only to compute the matrix-vector multiplications of the form $\mathbf{H}\mathbf{s}$ and $\mathbf{H}^H \mathbf{s}$. Thus, we can regard the matrix $\mathbf{H}$ as an operator and employ the FFT algorithm to reduce the complexity when calculating $\mathbf{H}\mathbf{s}$, as analyzed in Sec.~\ref{sec:complexity}.

\begin{algorithm}[t]
\caption{The iterative channel estimation and data detection method of DFT-s-OTFS}
\label{alg:iterative_ce_dd}
\KwIn{Received vectors $\mathbf{y}$ and $\mathbf{r}$, transmit pilot vector $\mathbf{x}_p$, the inverse of SNR $\lambda$.}
\KwOut{Detected data symbols $\hat{\mathbf{X}}_d$, estimated channel parameters $(\hat{\bm{\alpha}}, \hat{\bm{\tau}}, \hat{\bm{\nu}})$.}
\textit{Initialization:} $t = 0$\;
Solve \eqref{eq:coarse_ce} and obtain $(\hat{\bm{\alpha}}^{(0)}, \hat{\bm{\tau}}^{(0)}, \hat{\bm{\nu}}^{(0)})$\;
\Repeat{Stopping criteria}{
$t = t + 1$\;
Solve \eqref{eq:t_ls} using Algorithm~\ref{alg:cg} based on the received vector $\mathbf{r}$ and the estimated channel parameters $(\hat{\bm{\alpha}}^{(t-1)}, \hat{\bm{\tau}}^{(t-1)}, \hat{\bm{\nu}}^{(t-1)})$ and obtain $\hat{\mathbf{s}}^{(t)}$\;
Calculate $\hat{\mathbf{x}}^{(t)}$ according to \eqref{eq:estimated_Xd}, \eqref{eq:estimated_data} and \eqref{eq:estimated_x}\;
Solve \eqref{eq:iterative_ce} and obtain $(\hat{\bm{\alpha}}^{(t)}, \hat{\bm{\tau}}^{(t)}, \hat{\bm{\nu}}^{(t)})$\;
}
\Return{$\hat{\mathbf{X}}_d = \hat{\mathbf{X}}_d^{(t)}$, $(\hat{\bm{\alpha}}, \hat{\bm{\tau}}, \hat{\bm{\nu}}) = (\hat{\bm{\alpha}}^{(t)}, \hat{\bm{\tau}}^{(t)}, \hat{\bm{\nu}}^{(t)})$.}
\end{algorithm}

In this work, we develop a conjugate gradients (CG) based time-domain channel equalizer, which is an iterative method by using the efficient algorithms for the computation of $\mathbf{H}\mathbf{s}$ and reducing the computation complexity than the analytical solution of the LSP. The procedure of the CG method for the time-domain channel equalizer is described in Algorithm~\ref{alg:cg}.

With the estimated baseband time-domain transmit vector $\hat{\mathbf{s}}$ using Algorithm~\ref{alg:cg}, we are able to calculate the estimated information symbols as,
\begin{equation}\label{eq:estimated_Xd}
    \mathbf{X}_d' = \left(\text{vec}^{-1}\left(\hat{\mathbf{s}}\right) \mathbf{F}_N - \mathbf{X}_p\right) \mathbf{F}_N^H.
\end{equation}
Then the detected data symbols are obtained by solving the following demodulation problem, as
\begin{equation}\label{eq:estimated_data}
\hat{\mathbf{X}}_d = \arg \min_{\mathbf{X}_d \in \mathbb{A}^{M\times N}} \|\mathbf{X}_d - \mathbf{X}_d'\|_F^2.
\end{equation}
Thus, the estimated delay-Doppler domain transmit vector is updated as,
\begin{equation}\label{eq:estimated_x}
    \hat{\mathbf{x}} = \text{vec}\left(\hat{\mathbf{X}}_d \mathbf{F}_N\right) + \mathbf{x}_p.
\end{equation}
Based on the updated transmit vector, we are able to iterate channel estimation in \eqref{eq:iterative_ce}. The overall procedure of the iterative channel estimation and data detection for the joint passive sensing is illustrated in Algorithm~\ref{alg:iterative_ce_dd}. When the detected data symbols at $t$\textsuperscript{th} iteration are the same as those at $(t-1)$\textsuperscript{th} iteration, we can stop the iteration. In our simulations, we observe that the iterative channel estimation and data detection algorithm converge within 2 to 5 iterations. The main calculation steps lie in the CG based channel equalization, the complexity of which is $\mathcal{O}(MN \log(MN))$ with the usage of FFT algorithm to compute $\mathbf{H}\mathbf{s}$ and $\mathbf{H}^H \mathbf{s}$.

\section{Optimal Power Allocation Between Pilot and Data Symbols}\label{sec:power_allocation}

In the design of pilot placement, a non-zero pilot symbol is only located at one delay-Doppler grid point, while other pilot symbols are set as zero. Then the pilots are superimposed on the data symbols in the delay-Doppler domain. When the total power of the pilot and data symbols are constrained, it is necessary to optimally allocate the power of pilot. The tradeoff is elaborated as follows. On one hand, if we allocate more power to the non-zero pilot symbol, the channel estimation using superimposed pilots can become more accurate. However, in this case, less power is allocated to data symbols and may degrade the performance of data detection. On the other hand, if the pilot power is not sufficient, the pilot-aided channel estimation may become extremely inaccurate, which does not ensure the convergence of the iterative method. In this section, we optimize the power allocation between pilot and data to maximize the SINR by using the pilot- and data-aided channel estimation, which minimizes the BER.

Now we derive the effective SINR of the received data signal. The received vector in \eqref{eq:dd_io} can be written as,
\begin{equation}
    \mathbf{y} = \bm{\Omega}_p \bm{\alpha} + \bm{\Omega}_d \bm{\alpha} + \tilde{\mathbf{w}}
\end{equation}
where $\bm{\Omega}_p$ and $\bm{\Omega}_d$ are the concatenated matrices, given by
\begin{equation}
    \bm{\Omega}_p = \left[\begin{array}{cccc}
    \mathbf{\Gamma}_1 \mathbf{x}_p & \mathbf{\Gamma}_2 \mathbf{x}_p & \cdots & \mathbf{\Gamma}_P \mathbf{x}_p
\end{array}\right],
\end{equation}
\begin{equation}
    \bm{\Omega}_d = \left[\begin{array}{cccc}
    \mathbf{\Gamma}_1 \mathbf{x}_d^{\text{DD}} & \mathbf{\Gamma}_2 \mathbf{x}_d^{\text{DD}} & \cdots & \mathbf{\Gamma}_P \mathbf{x}_d^{\text{DD}}
\end{array}\right].
\end{equation}
The average path gain is defined as $\mathbb{E}\{\bm{\alpha}^H \bm{\alpha}\} = \sigma_h^2$. The estimated received data vector with pilot elimination using pilot- and data-aided channel estimation is derived as
\begin{equation}
\begin{split}
    \hat{\mathbf{y}}_d &= \mathbf{y} - \bm{\Omega}_p \hat{\bm{\alpha}} \\
    &= \bm{\Omega}_d \hat{\bm{\alpha}}+ (\bm{\Omega}_p + \bm{\Omega}_d) (\bm{\alpha} - \hat{\bm{\alpha}}) + \tilde{\mathbf{w}}.
\end{split}
\end{equation}
Thus, the SINR with the data-aided channel estimation is derived as,
\begin{equation}\label{eq:SINR}
\begin{split}
    \text{SINR} &= \frac{\mathbb{E}\left\{\left\|\bm{\Omega}_d \hat{\bm{\alpha}}\right\|^2\right\}}{\mathbb{E}\left\{\left\|(\bm{\Omega}_p + \bm{\Omega}_d) (\bm{\alpha} - \hat{\bm{\alpha}}) + \tilde{\mathbf{\Omega}}\right\|^2\right\}} \\
    &= \frac{\sigma_d^2 (\sigma_h^2 - \sigma_e^2)}{(\sigma_d^2 + \sigma_p^2) \sigma_e^2 + \sigma_w^2},
\end{split}
\end{equation}
where the error of the estimated channel coefficients $\sigma_e^2 = \mathbb{E}\left\{\left\|\bm{\alpha} - \hat{\bm{\alpha}}\right\|^2\right\}$, given by
\begin{equation}
    \sigma_e^2 = \frac{P(\sigma_h^2 \sigma_{xe}^2 + \sigma_w^2)}{MN(\sigma_p^2 + \sigma_d^2 - \sigma_{xe}^2)},
\end{equation}
where $\sigma_{xe}^2 = \mathbb{E}\left\{\left\|\mathbf{x}_d^{\text{DD}} - \hat{\mathbf{x}}_d^{\text{DD}}\right\|^2\right\}$ denotes the error of equalized data symbols with the estimated channel parameters using pilot-aided channel estimation, expressed as
\begin{equation}
    \sigma_{xe}^2 = \frac{P}{\frac{1}{\sigma_d^2} + \frac{\sigma_h^2 - \sigma_0^2}{\sigma_0^2 (\sigma_d^2 + \sigma_p^2) + \sigma_w^2}}.
\end{equation}
Here $\sigma_0^2$ denotes the error of the estimated channel coefficients $\mathbb{E}\left\{\left\|\bm{\alpha} - \hat{\bm{\alpha}}^{(0)}\right\|^2\right\}$ by using pilot-aided channel estimation, which is calculated as,
\begin{equation}
    \sigma_0^2 = \frac{P(\sigma_h^2 \sigma_d^2 + \sigma_w^2)}{MN \sigma_p^2}.
\end{equation}
The proof of the above derivations is relegated to Appendix~A.

With the derived effective SINR in \eqref{eq:SINR}, we can formulate the optimal power allocation scheme between pilot and data, given by
\begin{equation}\label{eq:optimal_sigma_p}
    \hat{\sigma}_p^2 = \arg\max_{\sigma_p^2 \in (0, 1)} \frac{\sigma_d^2 (\sigma_h^2 - \sigma_e^2)}{(\sigma_d^2 + \sigma_p^2) \sigma_e^2 + \sigma_w^2},
\end{equation}
where the constraint of average transmit power is imposed, $\sigma_p^2 + \sigma_d^2 = 1$, and the SNR of received signal becomes $\frac{\sigma_h^2}{\sigma_w^2}$.

\section{Simulation Results}\label{sec:simulation}

\begin{table}[!t]
    \centering
    \caption{Simulation Parameters}
    \label{tab:parameters}
    \begin{tabular}{|c|c|c|}
    \hline
    \hline
        \textbf{Notation} & \textbf{Definition} & \textbf{Value}  \\
         \hline
         \hline
        $f$ & Carrier frequency & 140 GHz, 0.3 THz \\
        \hline
        $\Delta f$ & Subcarrier spacing & 1.92 MHz \\
        \hline
        $T$ & Symbol duration & 1.04 $\mu s$\\
        \hline
        $M$ & Number of subcarriers & 64, 128 \\
        \hline
        $N$ & Number of symbols for one frame & 16, 32 \\
         \hline
         \hline
    \end{tabular}
\end{table}

In this section, we evaluate the performance of DFT-s-OTFS with superimposed pilots for THz integrated sensing and communication, including the PAPR, BER and sensing accuracy. We further compare it with other potential candidate waveforms, including OFDM, DFT-s-OFDM and OTFS. The simulation parameters are described in Table~\ref{tab:parameters}. The parameters are designed for THz and compatible with 5G numerology~\cite{Zaidi2016waveform}.



\subsection{PAPR and PA Efficiency}

In our preliminary work~\cite{wu2021dftsotfs}, we study the PAPR for discrete-time baseband signal. The PAPR of the discrete-time samples of DFT-s-OTFS frame transmit signal is defined as
\begin{equation}
    \text{PAPR} = \frac{P_\text{max}}{P_\text{avg}} = \frac{\max\left\{|s[m]|^{2}\right\}}{\frac{1}{M N} \sum_{m=0}^{MN-1}  |s[m]|^{2}},
\end{equation}
where $s[m]$ denotes the $m$\textsuperscript{th} element of $\mathbf{s}$.
Nevertheless, it is noticeably lower than the PAPR of the continuous-time baseband signal $s(t)$. We conduct $L$-times oversampling to obtain almost the same PAPR as $s(t)$. The oversampled signal can be generated by setting proper values of $\tau_i$ in \eqref{eq:s_tau}.

\begin{figure}
    \centering
    \includegraphics[width=0.45\textwidth]{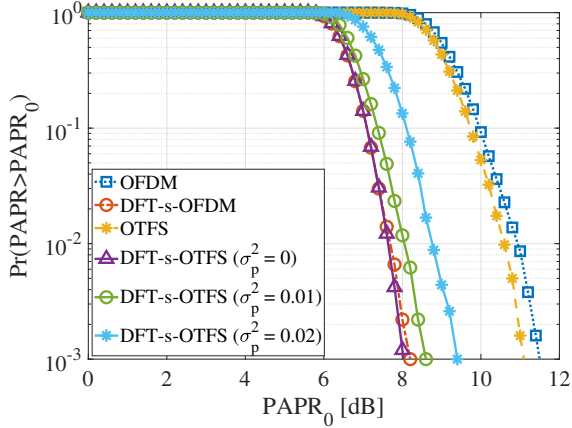}
    \caption{PAPR comparison for continuous-time baseband signal of OFDM, DFT-s-OFDM, OTFS and DFT-s-OTFS. The modulation scheme is 4-QAM. The subcarrier number $M = 64$ and the symbol number $N = 16$.}
    \label{fig:PAPR}
\end{figure}

\begin{figure}
    \centering
    \includegraphics[width=0.45\textwidth]{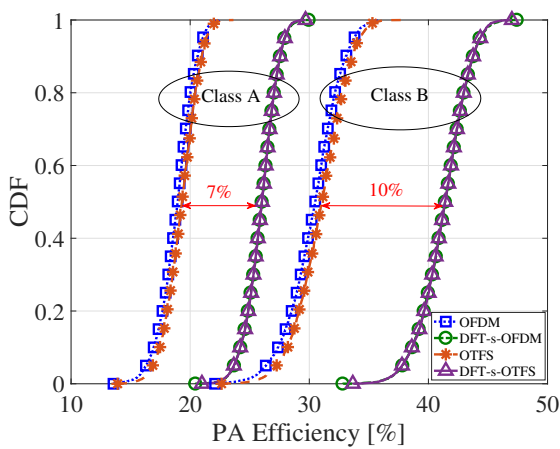}
    \caption{Comparison of theoretical efficiency limits for class A and class B power amplifiers when using different waveforms.}
    \label{fig:PA_efficiency}
\end{figure}

In Fig.~\ref{fig:PAPR}, we demonstrate that PAPR for DFT-s-OTFS continuous-time baseband signal can be approximately reduced by 3 dB, in contrast with OTFS. Meanwhile, when the DFT precoding size equals to $N$, DFT-s-OTFS has almost the same PAPR as DFT-s-OFDM. Moreover, the theoretical PA efficiency limits are calculated by $\eta = G \exp\left(- g \gamma_\text{dB}\right)$, where $\eta$ refers to the efficiency in \% and $\gamma_\text{dB}$ denotes the PAPR in dB. $G$ and $g$ have different values for class A and class B power amplifiers~\cite{miller1998peak}. We evaluate the PA efficiency limits for class A and class B power amplifiers in Fig.~\ref{fig:PA_efficiency}. For class A PA, the mean value of PA efficiency for DFT-s-OTFS can be improved by 7\%, compared with that for OTFS. For class B PA, the PA efficiency for DFT-s-OTFS is approximately 10\% higher than that for OTFS on average.

Next, we evaluate the influence of the superimposed pilots on the PAPR for DFT-s-OTFS. As shown in Fig.~\ref{fig:PAPR}, we set the signal power of the superimposed pilots in DFT-s-OTFS systems, $\sigma_p^2$, as different values, e.g., 0.01 and 0.02. We observe that as the signal power of pilots $\sigma_p^2$ increases, the PAPR of transmit signal for DFT-s-OTFS becomes higher, while it is still lower than that for OFDM and OTFS.

\subsection{Out-of-Band Emission}

\begin{figure}
    \centering
    \includegraphics[width=0.45\textwidth]{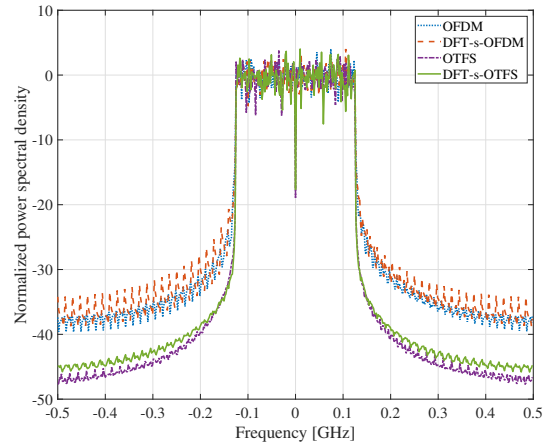}
    \caption{Comparison of out-of-band emission among OFDM, DFT-s-OFDM, OTFS and DFT-s-OTFS.}
    \label{fig:waveform_OOBE}
\end{figure}

A significant factor for the transmit waveform is the out-of-band power leakage. Large out-of-band power can incur adjacent channel interference (ACI). In this case, a guard band is required to reduce the effect of ACI, which causes the decrease of the spectral efficiency. Therefore, we investigate the out-of-band emission performance of OFDM, DFT-s-OFDM, OTFS and DFT-s-OTFS. The normalize power spectral density is compared in Fig.~\ref{fig:waveform_OOBE}. Here the number of used subcarriers is 128 and the number of symbols for a frame is 32. We demonstrate that the power leakage at both ends of the transmission band in OTFS and DFT-s-OTFS systems is approximately 10 dB lower than that in OFDM and DFT-s-OFDM systems.

\subsection{BER Performance}

\begin{figure}
    \centering
    \subfigure[The modulation scheme is 4-QAM.]{    \includegraphics[width=0.45\textwidth]{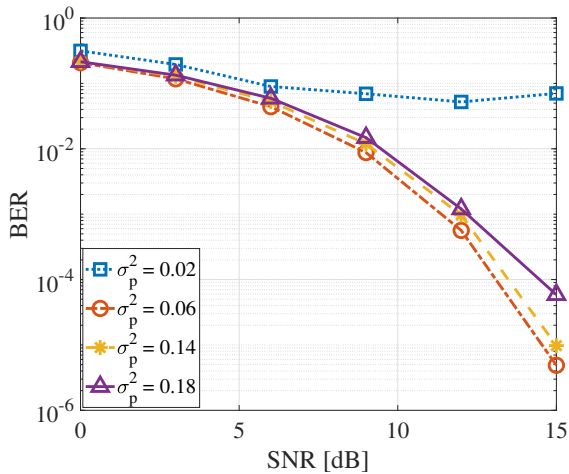}\label{fig:ber_4qam_140ghz}}
    \subfigure[The modulation scheme is 16-QAM.]{    \includegraphics[width=0.45\textwidth]{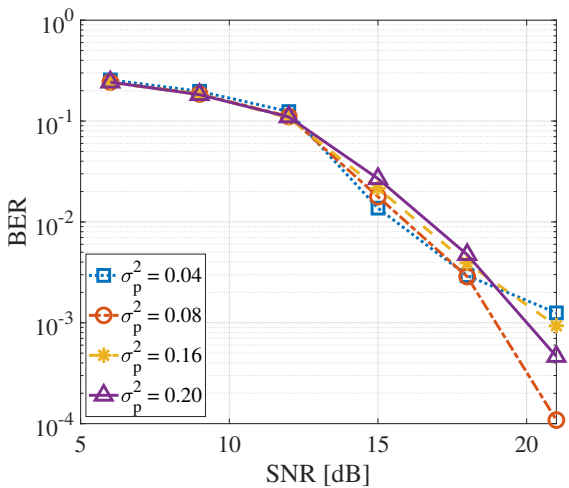}\label{fig:ber_16qam_140ghz}}
    \caption{BER performance of DFT-s-OTFS with superimposed pilots for different values of pilot power in a time-invariant channel at 140 GHz.}
    \label{fig:ber_140ghz}
\end{figure}

This section evaluates the BER performance of the proposed DFT-s-OTFS with the superimposed pilots. First, we consider an indoor time-invariant multi-path channel with a LoS path and two first-order reflected paths at 140 GHz~\cite{chen2021channel}. In Fig.~\ref{fig:ber_4qam_140ghz} and Fig.~\ref{fig:ber_16qam_140ghz}, we plot the BER of DFT-s-OTFS with superimposed pilots when using the iterative channel estimation data detection method in Sec.~\ref{sec:passive_sensing}. The modulation schemes are 4-QAM for Fig.~\ref{fig:ber_4qam_140ghz} and 16-QAM for Fig.~\ref{fig:ber_16qam_140ghz}, respectively. As shown in Fig.~\ref{fig:ber_4qam_140ghz}, we set four values for the average pilot symbol power $\sigma_p^2$, e.g., 0.02, 0.06, 0.14 and 0.18. At low SNR regime, these four cases have similar BER values. When the SNR is increased to 15 dB, we demonstrate that the BER performance for $\sigma_p^2 = 0.06$ is the best among the four values of $\sigma_p^2$. In addition, we learn that the BER for $\sigma_p^2 = 0.02$ is high at SNR of 15 dB, since the signal power allocated to pilots is small and causes inaccurate pilot-aided channel estimation and failure of iterative channel estimation and data detection.

\begin{figure}
    \centering
    \includegraphics[width=0.45\textwidth]{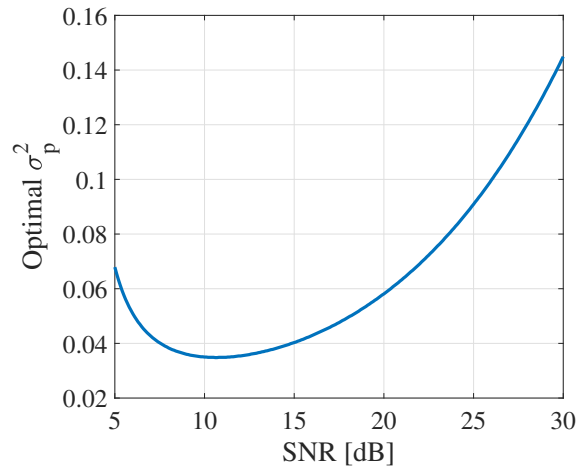}
    \caption{Optimal solution of power allocation between pilot and data versus SNR.}
    \label{fig:optimal_sigma_p_versus_SNR}
\end{figure}

In Fig.~\ref{fig:ber_16qam_140ghz}, since higher-order modulation scheme requires higher SNR to achieve good BER performance, we evaluate the BER from 6~dB to 21~dB, when the modulation scheme is 16-QAM. We observe that $\sigma_p^2 = 0.04$ has the best BER performance at SNR of 15 dB, while $\sigma_p^2 = 0.08$ achieves the lowest BER value at SNR of 21 dB. Meanwhile, we are able to use the optimal power allocation scheme between pilot and data in \eqref{eq:optimal_sigma_p} to calculate the optimal value of $\sigma_p^2$. The optimal solutions are $\hat{\sigma}_p^2 = 0.0403$ at SNR of 15 dB and $\hat{\sigma}_p^2 = 0.0633$ at SNR of 21 dB, which agree well with the simulation results in Fig.~\ref{fig:ber_4qam_140ghz} and Fig.~\ref{fig:ber_16qam_140ghz}. Moreover, by solving the optimization problem in \eqref{eq:optimal_sigma_p}, we are able to plot the optimal solution of allocated average pilot power versus SNR in Fig.~\ref{fig:optimal_sigma_p_versus_SNR}.
We learn that the optimal solution of $\sigma_p^2$ decreases with the increase of SNR at low SNR regime, while this optimal value increases as the SNR improves at high SNR regime.
At low SNRs, the performance of the pilot-aided channel estimation is noise-limited. Thus, when the noise power becomes stronger, more power should be allocated to the pilot for accurate channel estimation. At high SNRs, since the data signal is unknown at the communication receiver and regarded as interference when estimating the channel parameters based on the pilot, the channel estimation performance is primarily limited by the data signal power. In this case, we need to improve the pilot power with the increase of SNR. Therefore, it generates a saddle point at about 10 dB.

\begin{figure}
    \centering
    \includegraphics[width=0.45\textwidth]{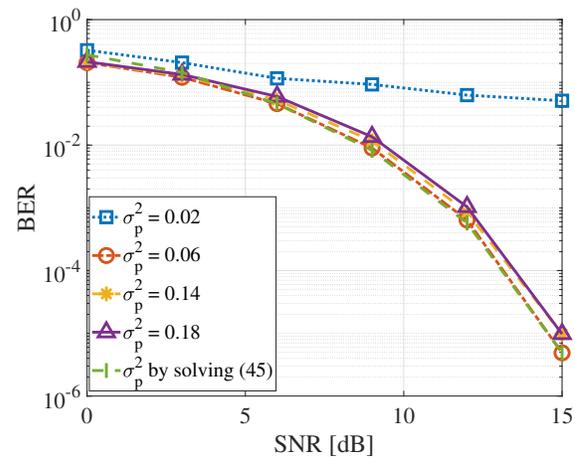}
    \caption{BER performance of DFT-s-OTFS with superimposed pilots for different values of pilot power in presence of strong Doppler effect at 0.3 THz. The modulation scheme is 4-QAM.}
    \label{fig:ber_4qam_300ghz}
\end{figure}

Next, we consider a doubly-selective channel at 0.3 THz, where the maximum velocity equals to 500 km/hr~\cite{wu2022dftsotfs-isac}. As shown in Fig.~\ref{fig:ber_4qam_300ghz}, we plot the BER performance of DFT-s-OTFS with superimposed pilots. With the proposed iterative channel estimation and data detection method, the simulation results indicate that the BER performance of DFT-s-OTFS is not degraded by the Doppler effect in such high-mobility channels. Thus, the proposed SI-DFT-s-OTFS system shows the same strong robustness to Doppler effects as OTFS. 
In addition, we learn that the BER performance with the optimal power allocation scheme is near optimal. At low SNRs, the pilot power by solving~\eqref{eq:optimal_sigma_p} performs worse than $\sigma_p^2 = 0.06, 0.14, 0.18$, since the derivation of \eqref{eq:SINR} invokes some approximations.

\subsection{Sensing Accuracy}

\begin{figure}
    \centering
    \includegraphics[width=0.45\textwidth]{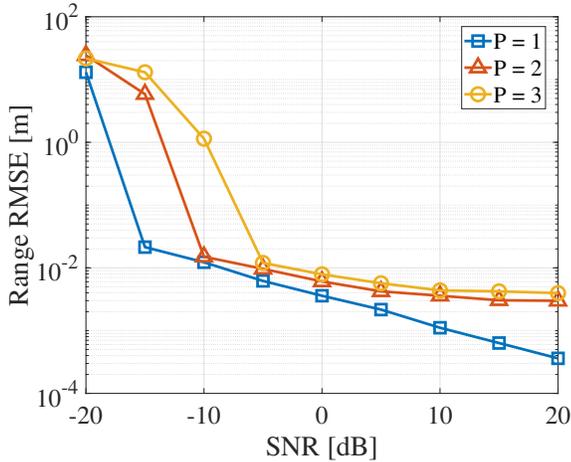}
    \caption{Range estimation accuracy of active sensing in DFT-s-OTFS by using the two-phase estimation method for different numbers of targets.}
    \label{fig:range_rmse}
\end{figure}

\begin{figure}
    \centering
    \includegraphics[width=0.45\textwidth]{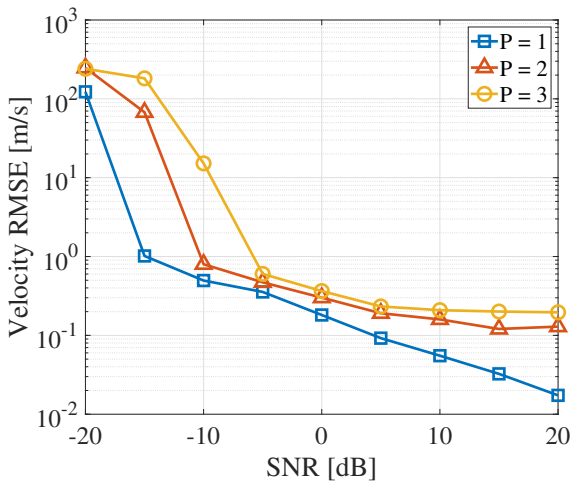}
    \caption{Velocity estimation accuracy of active sensing in DFT-s-OTFS by using the two-phase estimation method for different numbers of targets.}
    \label{fig:velocity_rmse}
\end{figure}

We further investigate the sensing accuracy of DFT-s-OTFS using the proposed two-phase estimation algorithm in Sec.~\ref{sec:active_sensing}. The performance metric is the root mean square error (RMSE).
The range RMSE metric is given by
$
    \text{RMSE}(r) = \sqrt{\mathbb{E}_{r\in \mathcal{D}} \left[\frac{1}{P}\sum_{i=1}^{P} \left(r_i - \hat{r}_i\right)^2 \right] },
$
where $r = [r_1, \cdots, r_P]$ and $\hat{r} = [\hat{r}_1, \cdots, \hat{r}_P]$ denote the real values and the estimated values of target range, respectively. $\mathcal{D}$ stands for the set of simulated samples. The velocity RMSE is given by
$
    \text{RMSE}(v) = \sqrt{\mathbb{E}_{v\in \mathcal{D}} \left[\frac{1}{P}\sum_{i=1}^{P} \left(v_i - \hat{v}_i\right)^2 \right] },
$
where $v = [v_1, \cdots, v_P]$ and $\hat{v} = [\hat{v}_1, \cdots, \hat{v}_P]$ represent the real values and the estimated values of target velocity, respectively.

For active sensing, we consider the parameters of three reference targets as, (10 m, 10 m/s), (30 m, 20 m/s), (50 m, 30 m/s). The used subcarriers is 128 and the symbol number is 32. In Fig.~\ref{fig:range_rmse}, we plot the range estimation accuracy of active sensing in DFT-s-OTFS by using the proposed two-phase estimation method for different numbers of targets, i.e., $P = 1, 2, 3$. In presence of single target, the RMSE of range estimation is less than 10\textsuperscript{-3} m above the SNR of 10 dB, i.e., millimeter-level range estimation accuracy. When there are multiple targets, i.e., $P = 2, 3$, the RMSE can achieve below 10\textsuperscript{-2} m. Meanwhile, we observe that there exists a SNR threshold for successful estimation. When the SNR is smaller than the threshold, the RMSE becomes very large. We learn that the threshold equals to -15, -10 and -5 dB for 1, 2 and 3 targets, respectively. This can be explained that since the received signal from one target may cause interference on estimating another target, the threshold becomes higher when the number of targets is increased.

\begin{figure}
    \centering
    \includegraphics[width=0.47\textwidth]{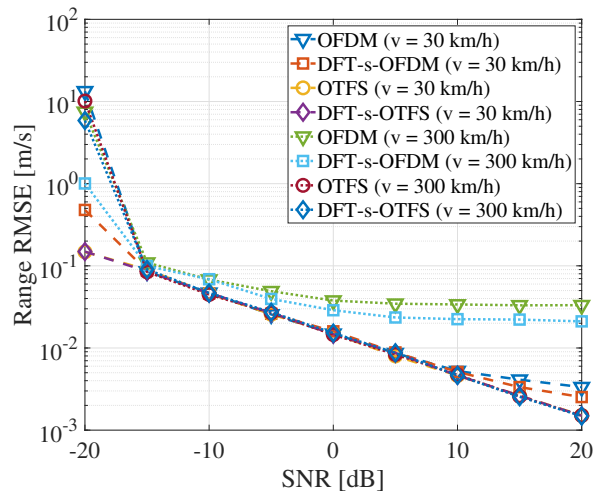}
    \caption{Range estimation accuracy comparison using OFDM, DFT-s-OFDM, OTFS and DFT-s-OTFS with different target velocity (30 km/h and 300 km/h).}
    \label{fig:range_compare}
\end{figure}

\begin{figure}
    \centering
    \includegraphics[width=0.47\textwidth]{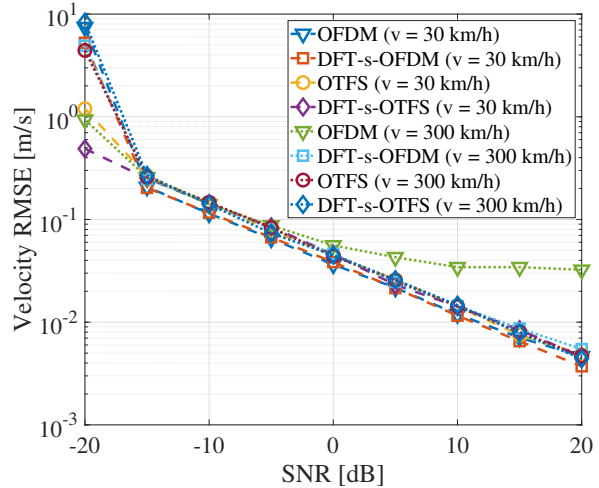}
    \caption{Velocity estimation accuracy comparison using OFDM, DFT-s-OFDM, OTFS and DFT-s-OTFS with different target velocity (30 km/h and 300 km/h).}
    \label{fig:velocity_compare}
\end{figure}

As shown in Fig.~\ref{fig:velocity_rmse}, we evaluate the RMSE of velocity estimation in DFT-s-OTFS system for active sensing. We demonstrate that the velocity estimation is able to achieve decimeter-per-second-level accuracy. While the increased number of targets results in larger estimation error, the simulation results indicate that the RMSE can approach 10\textsuperscript{-1} m/s for 2 and 3 targets. The effectiveness of the proposed two-phase estimation algorithm for the CDDS channels is validated by the accurate range and velocity estimation. The implementation complexity is reduced by the on-grid search in Phase I and the super-resolution estimation accuracy is guaranteed by the off-grid search in Phase II.
Moreover, we have evaluated the range and velocity estimation accuracy using OFDM, DFT-s-OFDM, OTFS and DFT-s-OTFS in Fig.~\ref{fig:range_compare} and Fig.~\ref{fig:velocity_compare}. The subcarrier spacing is set as 480 kHz. Different target velocity values are considered, including 30 km/h and 300 km/h. When the target velocity is 30 km/h, these waveforms have indistinguishable estimation accuracy. When the target velocity is increased to 300 km/h, the estimation accuracy of OFDM and DFT-s-OFDM is degraded, while the RMSE performance of OTFS and DFT-s-OTFS is not influenced.

\begin{figure}
    \centering
    \includegraphics[width=0.45\textwidth]{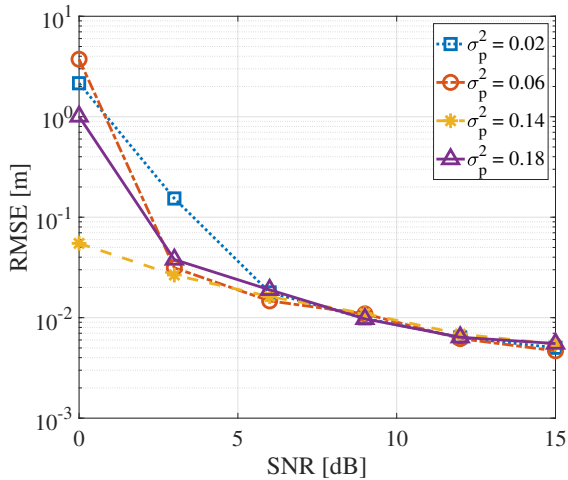}
    \caption{Range estimation accuracy of passive sensing in DFT-s-OTFS with superimposed pilots by using the iterative channel estimation and data detection method, $M$ = 128, $N$ = 32.}
    \label{fig:passive_sensing}
\end{figure}

Finally, we evaluate the performance of iterative channel estimation and data detection for passive sensing. We consider a THz channel with a LoS path and a NLoS path reflected by one sensing target at 0.3 THz.
The 4-QAM modulation scheme is used.
To compute the distance between the receiver and the target, we need to estimate the channel parameters with joint passive sensing and data detection. Suppose that the lengths of LoS path and the NLoS path are denoted by $r_{L}$ and $r_N$, respectively. Then the target distance is calculated as,
$
    r_s = \frac{r_N^2 - r_L^2}{2r_N - 2 r_L \cos\theta},
$
where $\theta$ represents the intersection angle of the LoS path and the NLoS path. The range estimation RMSE of passive sensing in DFT-s-OTFS with superimposed pilots by using the iterative channel estimation and data detection method is curved in Fig.~\ref{fig:passive_sensing}. At high SNR regime, the estimation error is below 10\textsuperscript{-2}, which has similar estimation accuracy as active sensing for multiple targets. At the SNR of 3 dB, the estimation accuracy for $\sigma_p^2 = 0.02$ is worse than other values of $\sigma_p^2$, since the allocated power to pilot is not high, which causes inaccurate coarse pilot-aided channel estimation. The simulation results show that the scheme of $\sigma_p^2 = 0.14$ performs well in terms of the passive sensing accuracy.

\section{Conclusion}\label{sec:conclusion}

In this paper, we have proposed a sensing integrated DFT-s-OTFS system framework for THz ISAC, which is applicable to two modes, i.e., active sensing, joint passive sensing and data detection.
We design a scheme of pilot placement in the delay-Doppler domain, in which the pilot symbols are superimposed onto the data symbols. Compared to embedded pilot scheme in OTFS systems, the design of superimposed pilots is able to improve the spectral efficiency without arranging a dedicated region for pilot placement. Furthermore, we develop an optimal power allocation scheme between pilot and data to optimize the performance of data recovery.
Based on the superimposed pilots, we have developed an iterative channel estimation and data detection method to estimate the channel parameters and recover data symbols.
To realize high-accuracy sensing, we derive a CDDS channel matrix and then propose a two-phase estimation algorithm for range and velocity parameter estimation.

With extensive simulations, the results indicate that DFT-s-OTFS can improve the PA efficiency by 7\% for class A PA and 10\% for class B PA compared with OTFS, since the PAPR of DFT-s-OTFS transmit signal is about 3 dB lower than OTFS. Meanwhile, DFT-s-OTFS with the proposed superimposed pilot-aided iterative channel estimation and data detection is robust to Doppler effects and the BER performance is not degraded in the high-mobility scenarios compared to time-invariant channels. Furthermore, by using the developed two-phase estimation algorithm, the proposed DFT-s-OTFS systems can achieve millimeter-level range estimation accuracy and decimeter-per-second-level velocity estimation accuracy.
In a nutshell, this work proposes DFT-s-OTFS as a promising candidate waveform for 6G sub-THz/THz communications, which is energy-efficient and has high robustness to Doppler shifts in the THz band. Meanwhile, it can achieve millimeter-level sensing accuracy with low complexity.

\section*{Appendix A}\label{sec:appendix_A}

When performing pilot-aided channel estimation, the mean square error of channel estimate is $\sigma_0^2 = \mathbb{E}\left\{\left\|\bm{\alpha} - \hat{\bm{\alpha}}_0\right\|^2\right\}$, derived as
\begin{equation}\label{eq:sigma_0_derivation}
\begin{split}
    \mathbb{E}\left\{\left\|\bm{\alpha} - \hat{\bm{\alpha}}^{(0)}\right\|^2\right\} &= \text{Tr}\left[\left(\bm{\Omega}_p^H\mathbf{C}_{w_d}^{-1}\bm{\Omega}_p\right)^{-1}\right] \\
    &= \left(\sigma_h^2 \sigma_d^2 + \sigma_w^2\right)\text{Tr}\left[\left(\bm{\Omega}_p^H\bm{\Omega}_p\right)^{-1}\right] \\
    &\approx \left(\sigma_h^2 \sigma_d^2 + \sigma_w^2\right) \frac{P^2}{\text{Tr}\left[\bm{\Omega}_p^H\bm{\Omega}_p\right]}\\
    &= \frac{P(\sigma_h^2 \sigma_d^2 + \sigma_w^2)}{MN \sigma_p^2},
\end{split}
\end{equation}
where $\mathbf{C}_{\mathbf{w}_d}$ denotes the covariance matrix of the data interference plus noise $\mathbf{w}_d = \bm{\Omega}_d\bm{\alpha} + \mathbf{\tilde{w}}$, given by $\mathbf{C}_{\mathbf{w}_d} = \mathbb{E}\left\{\mathbf{w}_d \mathbf{w}_d^H\right\} = \left(\sigma_h^2 \sigma_d^2 + \sigma_w^2\right) \mathbf{I}_{MN}$.

Next, with the estimated channel coefficients using pilots, the error of MMSE equalization is,
\begin{align}
    \sigma_{xe}^2 &= \mathbb{E}\left\{\left\|\mathbf{x}_d^{\text{DD}} - \hat{\mathbf{x}}_d^{\text{DD}}\right\|^2\right\} \notag\\
    &= \text{Tr}\left[\left(\left(\sum_{i=1}^P \hat{\alpha}_i^{(0)} \mathbf{\Gamma}_i\right)^H \mathbf{C}_{\mathbf{w}_0}^{-1} \left(\sum_{i=1}^P \hat{\alpha}_i^{(0)} \mathbf{\Gamma}_i\right) + \mathbf{C}_{\mathbf{x}_d}^{-1}\right)^{-1}\right] \notag\\
    &\approx \frac{P^2}{\text{Tr}\left[\left(\sum_{i=1}^P \hat{\alpha}_i^{(0)} \mathbf{\Gamma}_i\right)^H \mathbf{C}_{\mathbf{w}_0}^{-1} \left(\sum_{i=1}^P \hat{\alpha}_i^{(0)} \mathbf{\Gamma}_i\right) + \mathbf{C}_{\mathbf{x}_d}^{-1}\right]} \notag\\
    &= \frac{P}{\frac{1}{\sigma_d^2} + \frac{\sigma_h^2 - \sigma_0^2}{\sigma_0^2 (\sigma_d^2 + \sigma_p^2) + \sigma_w^2}},
\end{align}
where $\mathbf{w}_0 = \sum_{i=1}^P (\alpha_i - \alpha_i^{(0)}) \mathbf{\Gamma}_i (\mathbf{x}_d + \mathbf{x}_p) + \mathbf{\tilde{w}}$. Following the derivation in \eqref{eq:sigma_0_derivation}, we obtain the estimation error using pilot and data,
$\sigma_e^2 = \mathbb{E}\left\{\left\|\bm{\alpha} - \hat{\bm{\alpha}}\right\|^2\right\} 
    = \frac{P(\sigma_h^2 \sigma_{xe}^2 + \sigma_w^2)}{MN(\sigma_p^2 + \sigma_d^2 - \sigma_{xe}^2)}$.

\bibliographystyle{IEEEtran}
\bibliography{journal}

\begin{IEEEbiography}[{\includegraphics[width=1in,height=1.25in,clip,keepaspectratio]{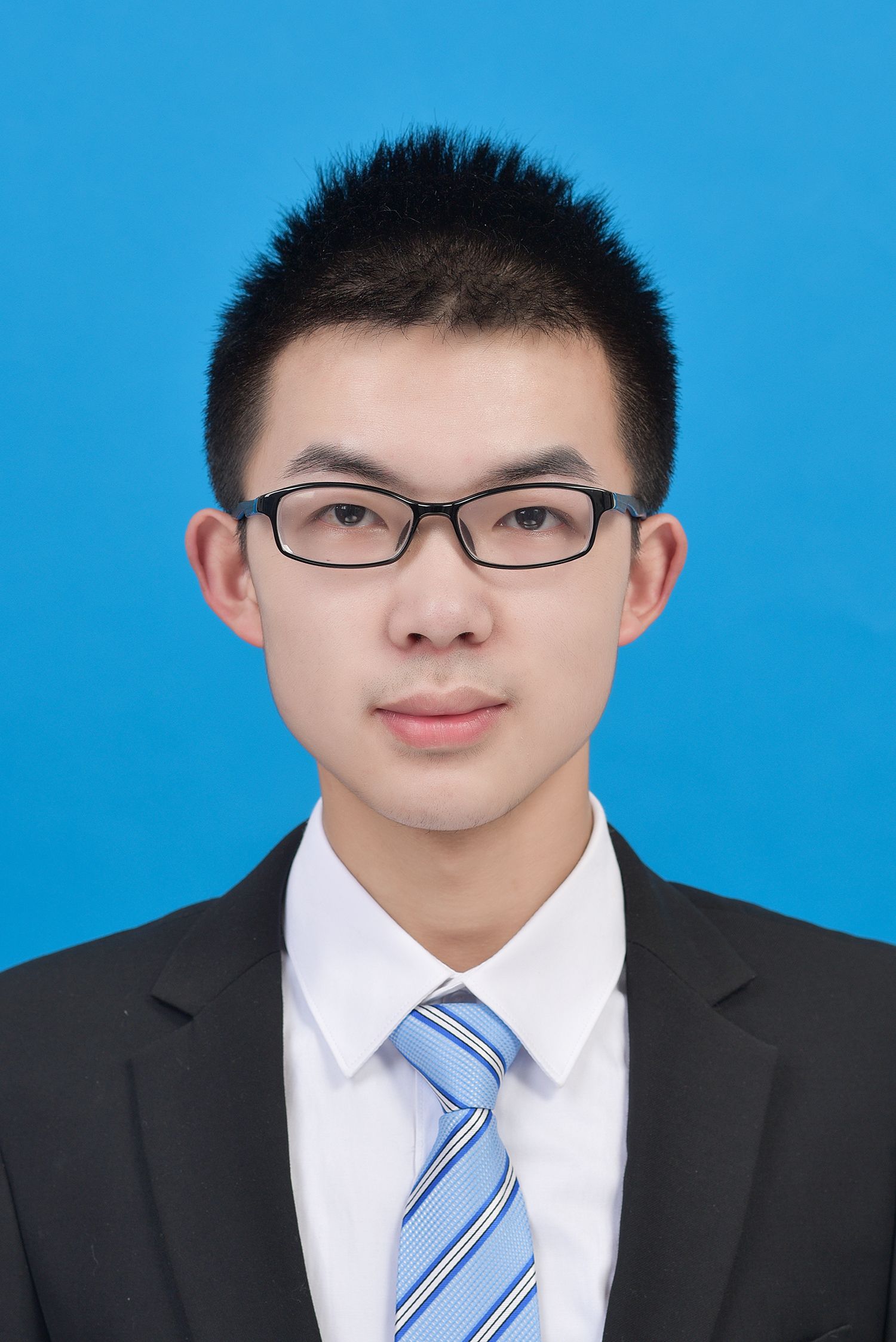}}]{Yongzhi Wu}
    (S'19) received B.E degree in Electronic and Information Engineering from Huazhong University of Science and Technology in 2019. Since 2019, he is pursuing Ph.D. degree in the Terahertz Wireless Communication Laboratory, Shanghai Jiao Tong University. His research interests include Terahertz communications, integrated sensing and communication.
    \end{IEEEbiography}

    \begin{IEEEbiography}[{\includegraphics[width=1in,height=1.25in,clip,keepaspectratio]{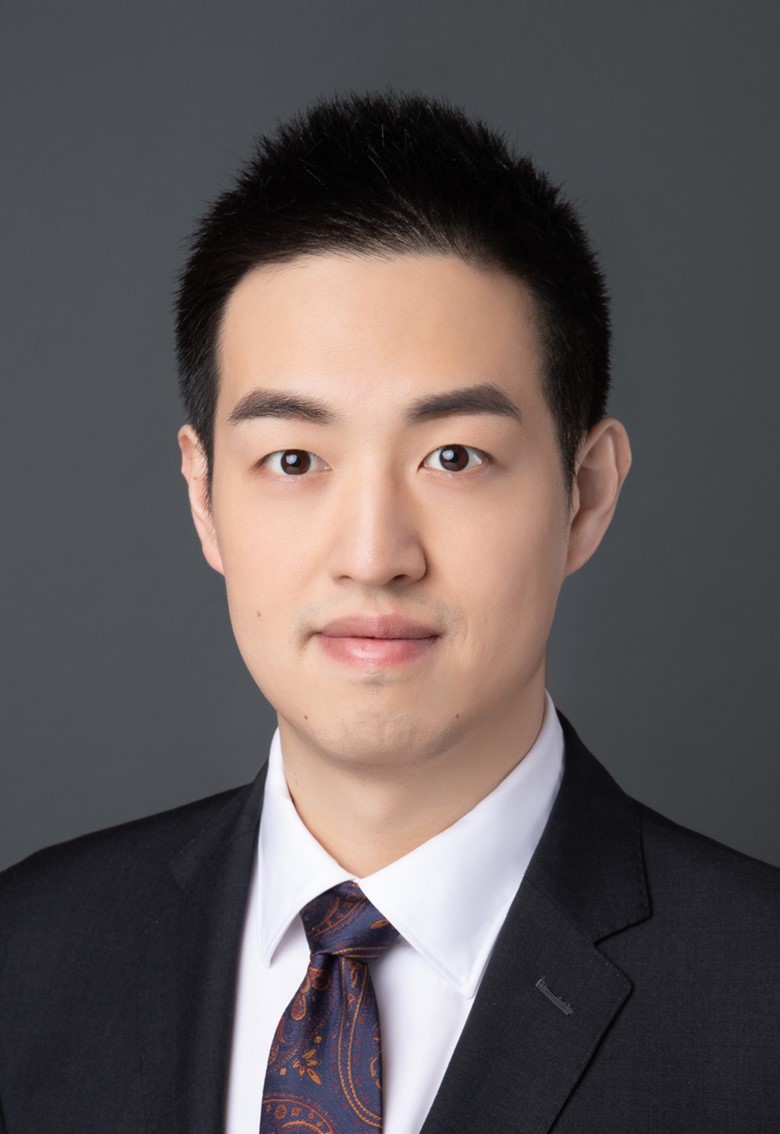}}]{Chong Han}
(M’16) received Ph.D. degree in Electrical and Computer Engineering from Georgia Institute of Technology, USA in 2016. He is currently a John Wu \& Jane Sun Endowed Associate Professor with University of Michigan-Shanghai Jiao Tong University (UM-SJTU) Joint Institute, Shanghai Jiao Tong University, China, and director of the Terahertz Wireless Communications (TWC) Laboratory. Since 2021, he is also affiliated with Department of Electronic Engineering and Cooperative Medianet Innovation Center (CMIC), Shanghai Jiao Tong University. He is the recipient of 2018 Elsevier NanoComNet (\textit{Nano Communication Network Journal}) Young Investigator Award, 2017 Shanghai Sailing Program, and 2018 Shanghai ChenGuang Program. He is a (guest) editor with \textsc{IEEE JSAC}, \textsc{IEEE JSTSP}, \textsc{IEEE Trans. Nanotechnology}, \textsc{IEEE Open Journal of Vehicular Technology} and etc. He is a TPC chair to organize multiple IEEE and ACM conferences and workshops, including GC’2023 SAC THz communications. He is a co-founder and vice-chair of IEEE ComSoc Special Interest Group (SIG) on Terahertz Communications, since 2021. His research interests include Terahertz and millimeter-wave communications. He is a member of the IEEE and ACM.
    \end{IEEEbiography}
    
    
    \begin{IEEEbiography}[{\includegraphics[width=1in,height=1.25in,clip,keepaspectratio]{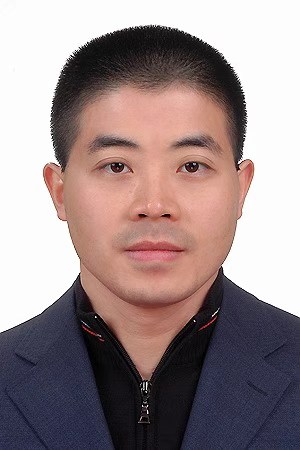}}]{Zhi Chen}
    (SM'16) received B. Eng, M. Eng., and Ph.D. degree in Electrical Engineering from University of Electronic Science and Technology of China (UESTC), in 1997, 2000, 2006, respectively. On April 2006, he joined the National Key Lab of Science and Technology on Communications (NCL), UESTC, and worked as a professor in this lab from August 2013. He was a visiting scholar at University of California, Riverside during 2010-2011. He is also the deputy director of Key Laboratory of Terahertz Technology, Ministry of Education. His current research interests include Terahertz communication, 5G mobile communications and tactile internet. 
    \end{IEEEbiography}

\ifCLASSOPTIONcaptionsoff
  \newpage
\fi

\end{document}